\documentclass[twocolumn,twoside,showkeys,preprintnumbers,amsmath,amssymb,secnumarabic, nofootinbib, nobibnotes,epl]{cbpf}
\pdfoutput=1
\usepackage[a4paper,dvips=false,pdftex=false,vtex=false]{geometry}
\usepackage{epsfig,epsf}
\usepackage{graphicx}
\usepackage{dcolumn}
\usepackage{bm}
\usepackage{fancyhdr}
\usepackage{pslatex}
\usepackage{pstricks}
\usepackage{graphicx,url}
\usepackage{float}
\usepackage[brazil]{babel}  
\usepackage[utf8]{inputenc}
\usepackage{setspace}
\usepackage{indentfirst}
\usepackage{enumerate}
\usepackage{tikz}
\usepackage{color}
\usepackage{array}
\usepackage{multirow}
\usepackage{booktabs}
\usepackage{makecell}
\usepackage{subcaption}
\usepackage{afterpage}
\usepackage{amsmath}
\usepackage{siunitx}
\usepackage{refcount}
\usepackage[nottoc]{tocbibind}
\usepackage{textcomp}
\usepackage{gensymb}
\usepackage{placeins}
\usepackage[normalem]{ulem}
\usepackage{eso-pic}
\usepackage{appendix}
\usepackage{varwidth}

\begin{document}
\linespread{1.0}

\noindent
{\footnotesize dx.doi.org/10.7437/NT2236-7640/?} \newline
{\footnotesize Notas T\'ecnicas, v. ?, n. ?, p. ?--?, 2021}
\\
\newline
\\
\\
\\\hspace*{-9cm}\rule{50em}{0.2ex}

\title{Classificação Morfológica de Galáxias no S-PLUS por Combinação de Redes Convolucionais\\
{\footnotesize \it Morphological Classification of Galaxies in S-PLUS using an Ensemble of Convolutional Networks}}

\author{N. M. Cardoso$^1$}
\email{nauxmac@gmail.com}
\author{G. B. O. Schwarz$^2$}
\email{gustavo.b.schwarz@gmail.com}
\author{L. O. Dias$^3$}
\author{C. R. Bom$^{3,4}$}
\author{L. Sodré Jr.$^5$}
\author{C. Mendes de Oliveira$^5$}

\affiliation{$^1$Escola Politécnica, Universidade de São Paulo, Av. Prof. Luciano Gualberto, 380, Butantã,  São Paulo -- SP, CEP 05508-010, Brasil}

\affiliation{$^2$Universidade Presbiteriana Mackenzie, Rua da Consolação, 930, Consolação, São Paulo -- SP, CEP 01302-907, Brasil}

\affiliation{$^3$Centro Brasileiro de Pesquisas Físicas, Rua Dr. Xavier Sigaud, 150, Urca,\\ Rio de Janeiro -- RJ, CEP 22290-180, Brasil}

\affiliation{$^4$
  Centro Federal de Educação Tecnológica Celso Suckow da Fonseca, Rodovia Mário Covas, lote J2, quadra J Distrito Industrial de Itaguaí, Itaguaí -- RJ. CEP: 23810-000, Brasil}

\affiliation{$^5$Departamento de Astronomia, Instituto de Astronomia, Geofísica e Ciências Atmosféricas da USP, Cidade Universitária, São Paulo -- SP, CEP 05508-900, Brasil}

\begin{abstract}
  \noindent
  {\bf Abstract:}  The universe is composed of galaxies that have diverse shapes. Once the structure of a galaxy is determined, it is possible to obtain important information about its formation and evolution. Morphologically classifying galaxies means cataloging them according to their visual appearance and the classification is linked to the physical properties of the galaxy. A morphological classification made through visual inspection is subject to biases introduced by subjective observations made by human volunteers. For this reason, systematic, objective and easily reproducible classification of galaxies has been gaining importance since the astronomer Edwin Hubble created his famous classification method. In this work, we combine accurate visual classifications of the Galaxy Zoo project with \emph {Deep Learning} methods. The goal is to find an efficient technique at human performance level classification, but in a systematic and automatic way, for classification of elliptical and spiral galaxies. For this, a neural network model was created through an Ensemble of four other convolutional models, allowing a greater accuracy in the classification than what would be obtained with any one individual. Details of the individual models and improvements made are also described. The present work is entirely based on the analysis of images (not parameter tables) from DR1 (www.datalab.noao.edu) of the Southern Photometric Local Universe Survey (S-PLUS). In terms of classification, we achieved, with the Ensemble, an accuracy of $\approx 99 \%$ in the test sample (using pre-trained networks).

  \vspace*{3mm}

  \vspace*{3mm}
  \noindent
  {\bf Resumo:} O universo é composto de galáxias que apresentam variadas formas. Uma vez determinada a estrutura de uma galáxia, é possível obter informações importantes desde sua formação até sua evolução. A classificação morfológica é a catalogação de galáxias de acordo com a sua aparência visual e a classificação está ligada com as propriedades físicas da galáxia. Uma classificação morfológica feita através de inspeção visual está sujeita a um viés causado pela subjetividade da observação humana. Por isso, a classificação sistemática, objetiva e facilmente reproduzível de galáxias vem ganhando importância desde quando o astrônomo Edwin Hubble criou seu famoso método de classificação. Neste trabalho, nós combinamos classificações visuais acuradas do projeto Galaxy Zoo com métodos de \emph{Deep Learning}. O objetivo é encontrar uma técnica eficiente que consiga simular a classificação visual humana, mas de forma sistematizada e automática, para classificação de galáxias elípticas e espirais. Para isto, um modelo de rede neural foi criado através de um Ensemble de outros quatro modelos convolucionais, possibilitando uma maior acurácia na classificação do que o que seria obtido com qualquer um individualmente. Detalhes dos modelos individuais e melhorias feitas nestes também são descritas.   O presente trabalho é totalmente baseado na análise de imagens (não tabelas de parâmetros) do DR1 (www.datalab.noao.edu) do Southern Photometric Local Universe Survey (S-PLUS). Em termos de classificação, alcançamos, com o Ensemble, uma precisão de  $\approx 99 \%$ na amostra de teste (usando redes pre-treinadas).

  \vspace*{3mm}
  \noindent
  {\bf Palavras chave:} Morfologia de Galáxias, Classificação de Galáxias, Visão Computacional, Aprendizagem Profunda, Redes Neurais Convolucionais.
\end{abstract}

\maketitle
\setcounter{page}{1}

\section{Introdução}
\label{sec:intro}
Classificação morfológica é a categorização das galáxias conforme sua forma. Quando esta classificação é baseada na inspeção visual das imagens, elementos subjetivos são agregados. Em 1926, o astrônomo Edwin Hubble, na tentativa de relacionar as formas das galáxias com sua origem e evolução, criou um método hoje conhecido como \emph{Hubble Sequence} ou \emph{Tunning Fork} \cite{hubble1926, fortson2012}, que é uma tentativa de atribuir classes discretas às galáxias, de acordo com suas formas. Esta classificação, com algumas pequenas modificações e adições, ainda é usada até hoje.  No \emph{Tunning Fork}, as galáxias são classificadas como elípticas, espirais ou lenticulares, mas as formas predominantes de grandes galáxias na natureza são elípticas e espirais \cite{fortson2012}, pois acredita-se que a classe das lenticulares seja uma classe de transição. Lenticulares são muitas vezes classificadas como galáxias elípticas (mais comumente) ou espirais (menos comum). Com isto, foram criadas as classes "early-type", contendo as elípticas e lenticulares e "late-type", contendo as espirais e outras galáxias de tipo mais tardio ainda, chamadas de irregulares, que só foram incluídas no sistema de classificação muitos anos mais tarde.

O final do século 20 conheceu uma revolução na maneira de se estudar galáxias na Astronomia quando os primeiros mapeamentos de grandes áreas do céu começaram a ser feitos. O mapeamento que mais impactou a Astronomia nas últimas décadas foi o chamado SDSS\footnote{SDSS: Sloan Digital Sky Survey -- \url{https://www.sdss.org}.}.

Um programa que envolveu o SDSS e milhões de cidadãos comuns (chamado, em inglês, de projeto \emph{citizen science}) foi o chamado GalaxyZoo\footnote{\url{https://galaxyzoo.org}}, um projeto realizado em sua maioria por cidadãos sem vínculo acadêmico, que contribuíram com suas observações para a classificação de um grande número de galáxias do SDSS. A segunda liberação de dados do GalaxyZoo possui um catálogo com classificações morfológicas de trezentas mil galáxias, revisadas segundo o método de Hart et al. \cite{hart2016}. Uma subamostra destes dados, que coincide com o chamado \emph{Stripe-82}\footnote{Este é um campo equatorial do céu de 336 graus$^{2}$, que cobre a região com ascensão reta das 20:00h às 4:00h e declinação de -1,26$^{\circ}$ a +1,26$^{\circ}$}, foi utilizada neste trabalho como \emph{true table} na classificação de galáxias elípticas e espirais.

Com o avanço dos levantamentos (\emph{surveys}) digitais e conseguente aumento da quantidade de dados coletados, se torna crucial o desenvolvimento de métodos rápidos e automatizados para a classificação morfológica de galáxias sem a perda da acurácia da tradicional classificação visual \cite{yamauchi2005}. O uso de aprendizado de máquina e, mais recentemente \emph{Deep Learning} tem mostrado resultados relevantes para problemas de classificação em diversos problemas nas áreas de visão computacional e astronomia, dentre outras.

O \emph{Deep Learning} \cite{Goodfellow2016} é uma segmento específico dentro da área de aprendizado de máquina e, por conseguinte, da área de Inteligência Artificial. Consiste no desenvolvimento de redes neurais artificiais que são combinadas em um número significativamente maior do que as redes neurais tradicionais. Este tipo de técnica se transformou no estado-da-arte do reconhecimento de padrões em imagens devido a um tipo específico de rede neural conhecida como convolucional.
As redes neurais convolucionais ou CNNs da sigla em inglês \textit{Convolutional Neural Networks} \cite{lecun2015deep}, são inspiradas e propostas com certa analogia ao processamento das imagens realizadas no córtex visual de mamíferos. O processo começa quando um estímulo visual alcança a retina e equivale a um sinal que atravessa regiões específicas do cérebro. Essas regiões são responsáveis pelo reconhecimento de cada uma dessas características correspondentes \cite{karpathy2016convolutional}.
Os neurônios biológicos das primeiras regiões respondem pela identificação de formatos geométricos primários, enquanto neurônios das camadas finais têm a função de detectar características mais complexas, formadas pelas formas simples anteriormente reconhecidas \cite{karpathy2016convolutional,vedaldi2015matconvnet}. Características com padrões muito específicos do objeto são estabelecidas depois que o procedimento se repete.
De forma análoga, a CNN decompõe a tarefa de reconhecimento de um objeto em subtarefas. Para isso, durante a aprendizagem, a CNN divide a tarefa em subníveis de representação das características, posteriormente aprendendo a reconhecer novas amostras da mesma classe  \cite{lecun2015deep,vedaldi2015matconvnet}.
Desta forma, as CNNs são capazes de predizer características complexas sem a necessidade de um pré processamento e são invariantes â escala e à rotação dos dados, o que torna essencial a classificação em imagens.

Este trabalho utilizou dados do S-PLUS para classificar imagens. O S-PLUS \cite{oliveira2019} é um levantamento de galáxias do Universo Local, liderado por brasileiros, feito com um telescópio de 0.8m e com uma câmera de grande campo, localizado no Chile. A parte do mapeamento que cobre a região do chamado \emph{Stripe-82} é uma área de grande interesse dado que é coberta por diversos projetos, permitindo assim comparações e análises complementares. O S-PLUS cobriu a região com medidas de fluxo (magnitudes) em 12 bandas para três milhões de fontes (liberadas para a comunidade internacional no DR1, \cite{oliveira2019}).

\begin{figure*}[!ht]
  \centering
  \includegraphics[width=\linewidth]{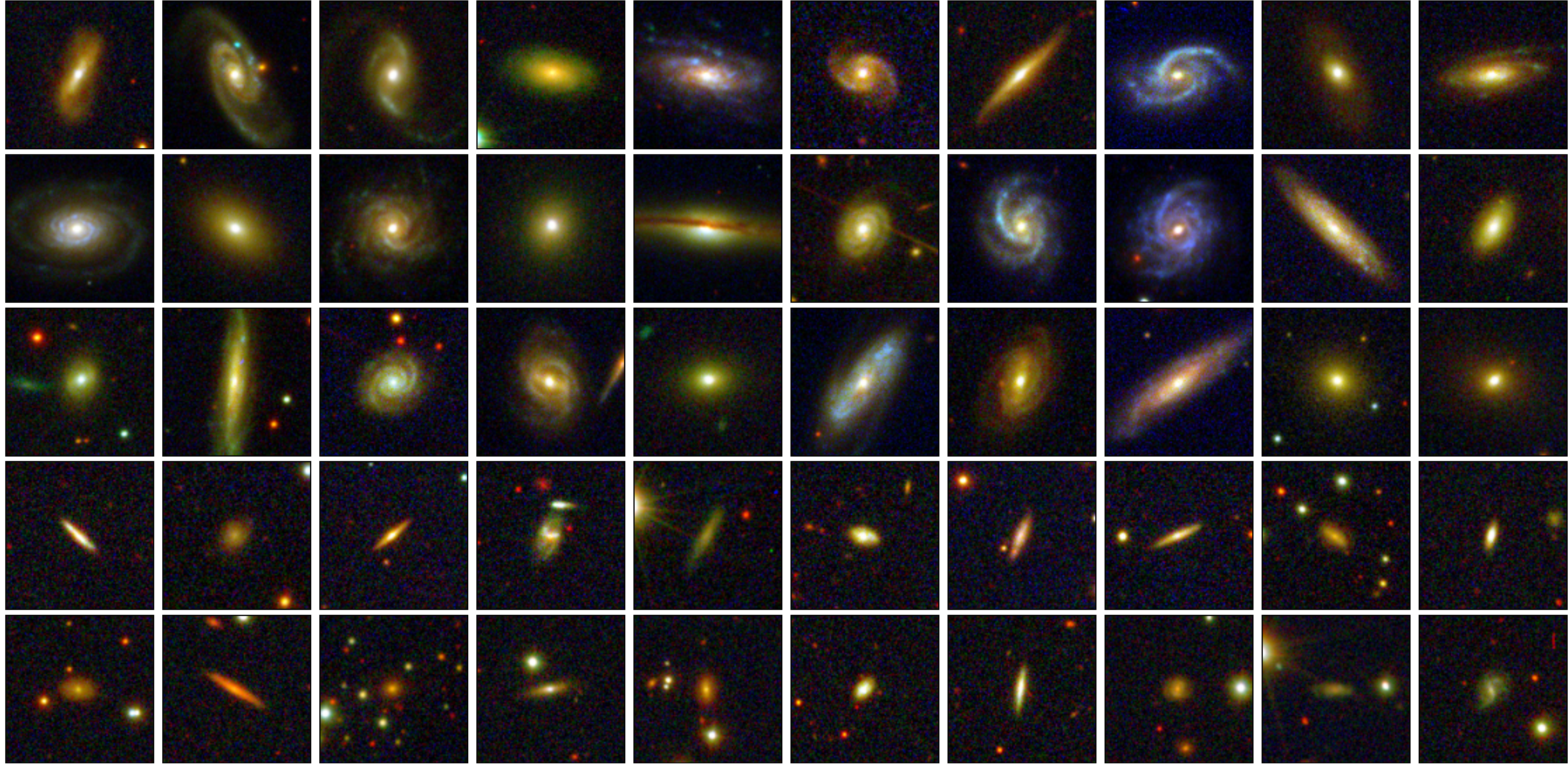}
  \caption{Exemplos de galáxias utilizadas nos conjuntos de treinamento, validação e teste, coloridas usando o método descrito na Seção \ref{section:preparacao}. As primeiras três linhas mostram galáxias com $r_{auto} < 17$ e as últimas duas mostram galáxias no intevalo de magnitude $17 < r_{auto} < 17.5$. Note a diferença entre as imagens de diferentes magnitudes, na Seção \ref{section:resultados} será mostrado que esta diferença tem impacto na performance do modelo.}
  \label{fig:galaxy_grid}
\end{figure*}

Neste artigo, a Seção 2 versa sobre a preparação dos dados. A Seção 3 apresenta as seis redes com suas performances e seus respectivos hiperparâmetros. Quatro das redes com melhor performance são então escolhidas para a fase seguinte do trabalho, para construção de um meta-modelo, com os resultados apresentados na Seção 4. Mostramos que as características das galáxias sem prévia classificação, classificadas com nosso método, apresentam as cores esperadas, indicativo de uma classificação robusta. Finalmente, a Seção 5 apresenta a discussão dos resultados. Comparações são feitas entre os nossos resultados e os obtidos através de classificações feitas por outros autores.

\section{Conjunto de Dados}

Para este trabalho queremos desenvolver uma técnica eficiente e automatizada para a  classificação morfológica de galáxias usando \emph{Deep Learning}. Para isso, primeiramente serão apresentados os dados utilizados e como os preparamos.

\subsection{Aquisição dos dados}

A imagem da galáxia, com sua respectiva classificação morfológica, é elemento crucial para se fazer o treinamento supervisionado de nossa Rede Neural Artificial. Para treinar a rede, de modo que esta aprenda a classificar galáxias através das imagens, é necessário fazê-la aprender as formas e padrões das galáxias. No nosso caso, utilizamos uma grande amostra de imagens de galáxias já classificadas visualmente por humanos.

Os dados aqui utilizados são as imagens das galáxias do levantamento S-PLUS e as classificações morfológicas do GalaxyZoo, que separam as galáxias entre espiral e elíptica. A associação destes dois conjuntos de dados é feita pela correlação das coordenadas do objeto no espaço. Ademais, apenas galáxias no S-PLUS com magnitudes $r_{auto}$ menores que 17.5 foram utilizadas. O valor de $r_{auto}$, que é dado em uma das colunas do catálogo do S-PLUS, representa aproximadamente o fluxo total de uma dada galáxia. Mais adiante, na Seção \ref{section:resultados}, faremos uma análise usando dois conjuntos de dados, um com galáxias com $r_{auto} < 17$ e outro com galáxias com $r_{auto} < 17.5$, para mostrar a importância do limite de magnitude ($r_{auto}$) nos resultados. O número de galáxias em cada um destes conjuntos é dado na tabela \ref{tab:conjuntos}.  As imagens do S-PLUS foram obtidas através do banco de dados do projeto\footnote{\url{https://splus.cloud}}. A Figura \ref{fig:galaxy_grid} mostra exemplos d imagens RGB do S-PLUS utilizadas neste trabalho.

\subsection{Divisão do conjunto de dados}
\label{section:divisao_conjunto_de_dados}

Todos os modelos usam os mesmos subconjuntos de dados para garantir que o desempenho de classificação do modelo não seja enviesado pela escolha dos lotes.

A distribuição de galáxias para cada subconjunto foi de 81\% para treinamento, 9\% para validação e 10\% para teste. A proporção de galáxias elípticas e espirais entre os subconjutos de treinamento, validação e teste é a mesma. Os conjuntos são constituídos por, aproximadamente, 68\% de galáxias espirais e 32\% de galáxias elípticas. A tabela \ref{tab:conjuntos} mostra a quantidade de galáxias em cada  subconjunto.

\begin{table}[!h]
  \centering
  \caption{Quantidade de galáxias em cada conjunto de dados.}
  \label{tab:conjuntos}%
  \begin{tabular}{*{5}{c}}
    \toprule
    $r_{auto}$ & Treinamento & Validação & Teste & Total \\
    \midrule
    $< 17$     & 2231        & 248       & 276   & 2757  \\
    $< 17.5$   & 3349        & 373       & 414   & 4136  \\
    \bottomrule
  \end{tabular}
\end{table}

\subsection{Comparação dos conjuntos de dados}

\begin{figure}[!ht]
  \centering
  \includegraphics[width=\linewidth]{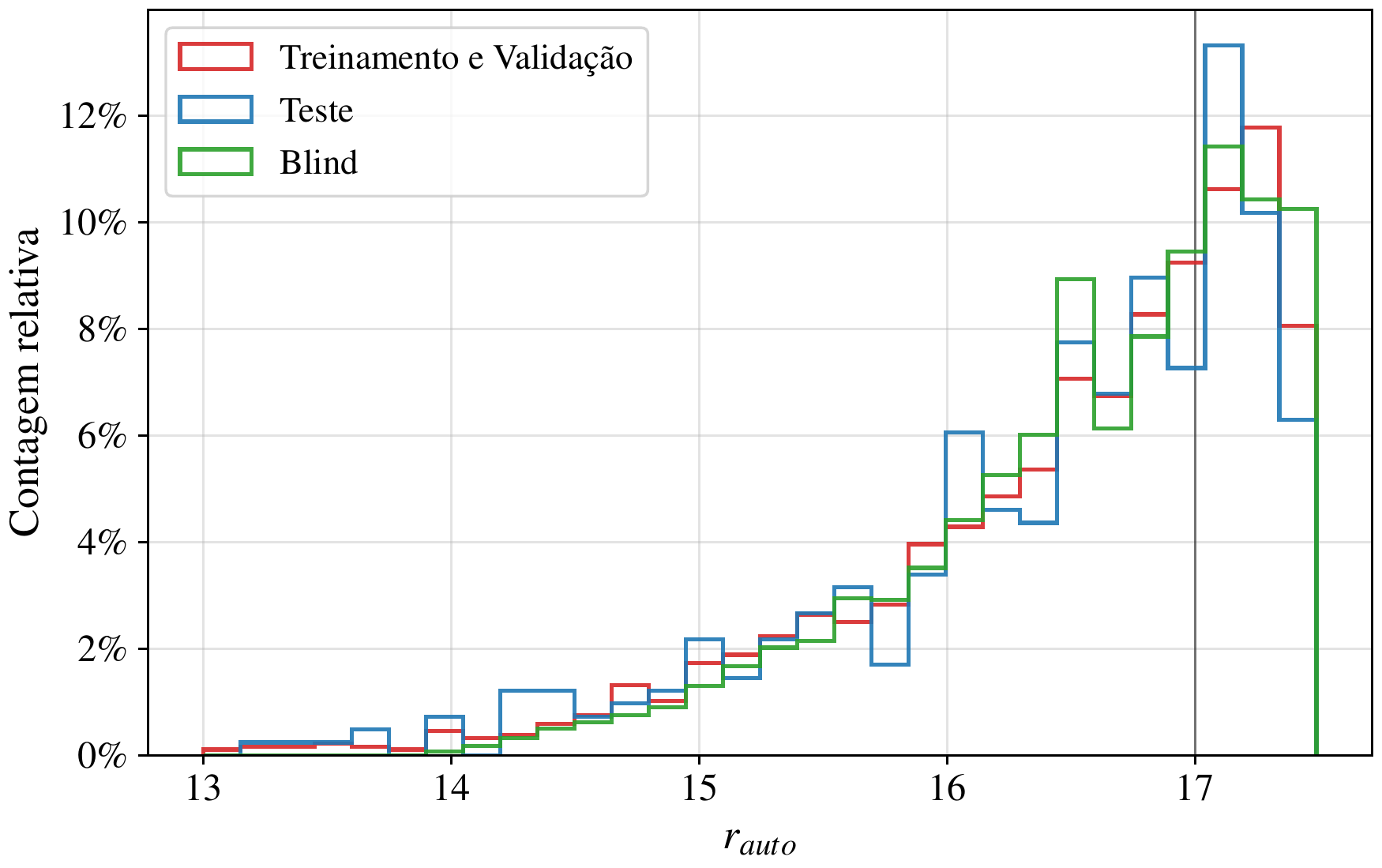}
  \caption{Histograma com a contagem relativa dos valores de $r_{auto}$ para as amostras de treinamento e validação, teste e blind.}
  \label{fig:dist-rauto}
\end{figure}

\begin{figure}[!ht]
  \centering
  \includegraphics[width=\linewidth]{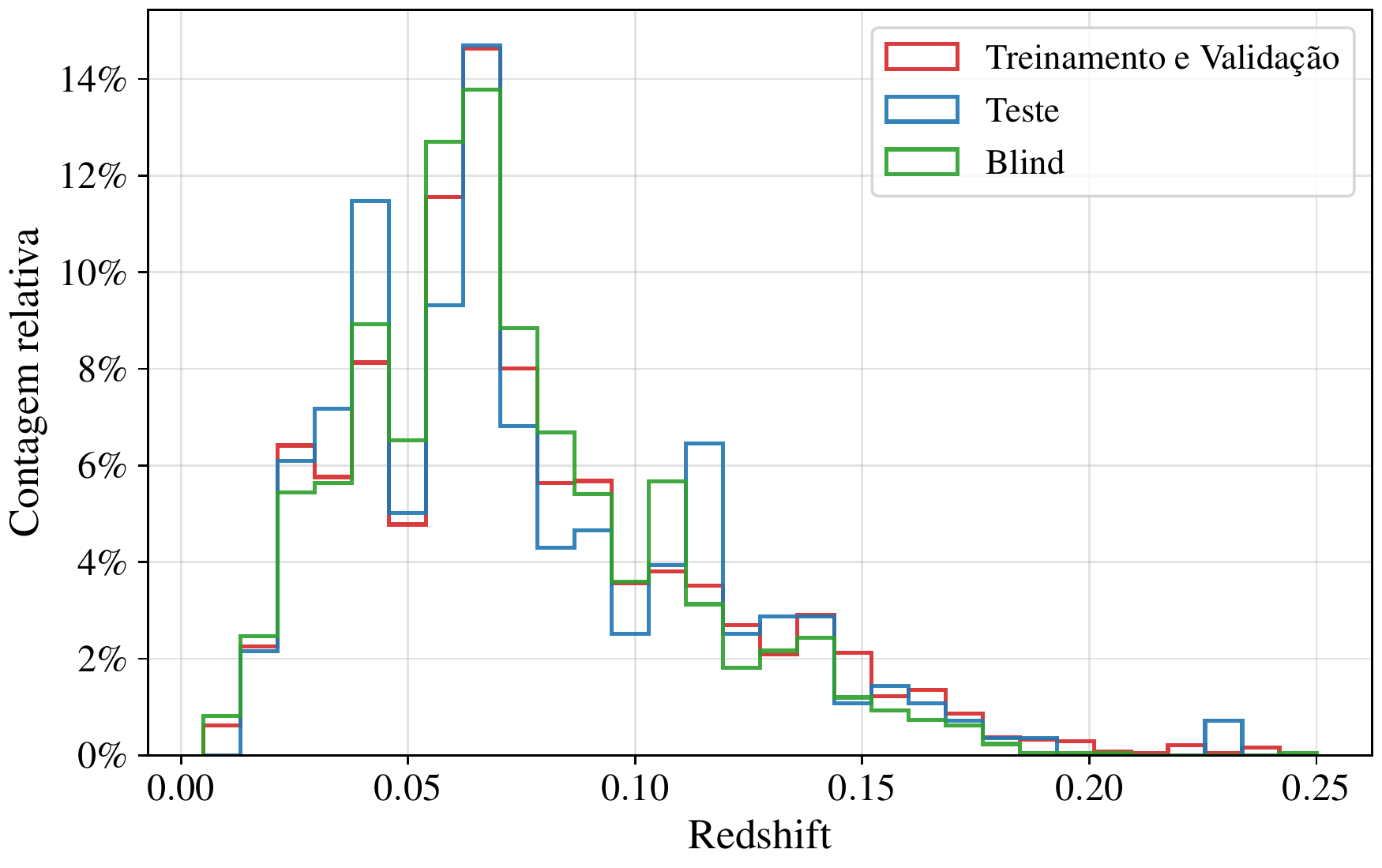}
  \caption{Histograma com a contagem relativa dos valores de redshift para as amostras de treinamento e validação, teste e blind para $r_{auto} < 17$.}
  \label{fig:redshift-17}
\end{figure}

\begin{figure}[!ht]
  \centering
  \includegraphics[width=\linewidth]{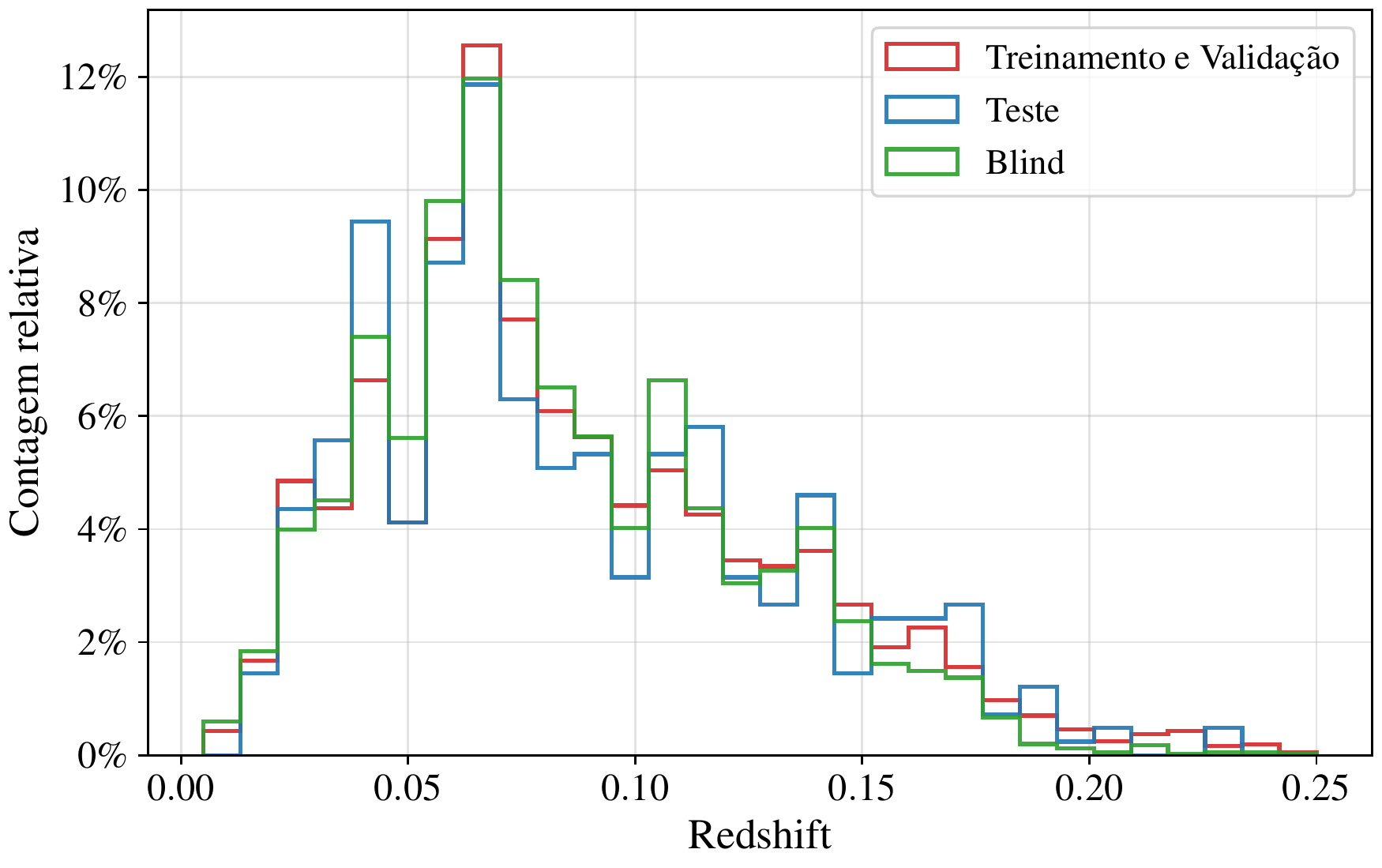}
  \caption{Histograma com a contagem relativa dos valores de redshift para as amostras de treinamento e validação, teste e blind para $r_{auto} < 17.5$.}
  \label{fig:redshift-17.5}
\end{figure}

Para o desenvolvimento do trabalho, dividimos as amostras em subconjuntos distintos, com o objetivo final de fazer uma classificação de galáxias elípticas e espirais numa amostra ainda não classificada, denominada amostra \emph{blind}. Nesta seção, mostramos que as amostras de galáxias utilizadas nos conjuntos de treinamento, validação, teste e \emph{blind} têm distribuições equilibradas de medidas de brilho, determinadas através das magnitudes na banda r (que é a banda com maior sinal/ruído) e redshifts (desvios para o vermelho, que são proporcionais às distâncias dos objetos). Isto é importante para que a comparação dos diagramas cor-cor que serão mostrados na Figura \ref{fig:color-color}, na Seção \ref{section:resultados} faça sentido. A Figura \ref{fig:dist-rauto} mostra o histograma de magnitude das galáxias para valores de $r_{auto}$ entre 13 e 17.5. Uma linha em 17 mostra que galáxias à direita estão presentes somente no conjunto $r_{auto} < 17.5$. As Figuras \ref{fig:redshift-17} e \ref{fig:redshift-17.5} mostram as distribuições de redshift para os conjuntos $r_{auto} < 17$ e $r_{auto} < 17.5$, respectivamente.

\subsection{Conjunto de dados desbalanceado}

Como visto na Seção \ref{section:divisao_conjunto_de_dados}, aproximadamente 68\% das galáxias do conjunto de dados são espirais. Contudo, o desempenho dos algorítmos de \emph{machine learning} são afetados negativamente pela quantidade desproporcional de objetos entres as classes. Algumas técnicas testadas para melhorar o desempenho da rede são listadas abaixo.

\begin{description}
  \item[Subamostragem aleatória] \hfill \\
        A subamostragem aleatória, \emph{random undersampling}, é a retirada aleatória de objetos do conjunto de trainamento pertencentes à classe com maior quantidade de elementos até que a proporção de objetos entre as classes fique equilibrada ($\approx 1:1$).

  \item[Sobreamostragem aleatória] \hfill \\
        Ao contrário da subamostragem, a sobreamostragem aleatória, \emph{random oversampling}, é a replicação de elementos da classe em minoria até que a proproção de objetos entre as classes fique equilibrada.

  \item[Ponderamento das classes] \hfill \\
        Ao contrário das técnicas anteriores, o ponderamento das classes (\emph{class weight}) não é uma técnica de reamostragem. Ela consiste na atribuição de pesos a cada classe, proporcionais a sua quantidade de elementos.
        O peso $w_i$ da i-ésima classe tem o valor dado pela equação \eqref{eq:pesos}, baseada na heurística apresentada em \cite{KinZen01}.
        \begin{equation}
          w_i = \frac{Q}{N \times C_i}
          \label{eq:pesos}
        \end{equation}
        onde, $Q$ é a quantidade total de objetos, $N$ é o número de classes e $C_i$ é o número de objetos da i-ésima classe.
\end{description}

\section{Métodos de Deep Learning}

\begin{figure*}[!ht]
  \centering
  \includegraphics[width=\linewidth]{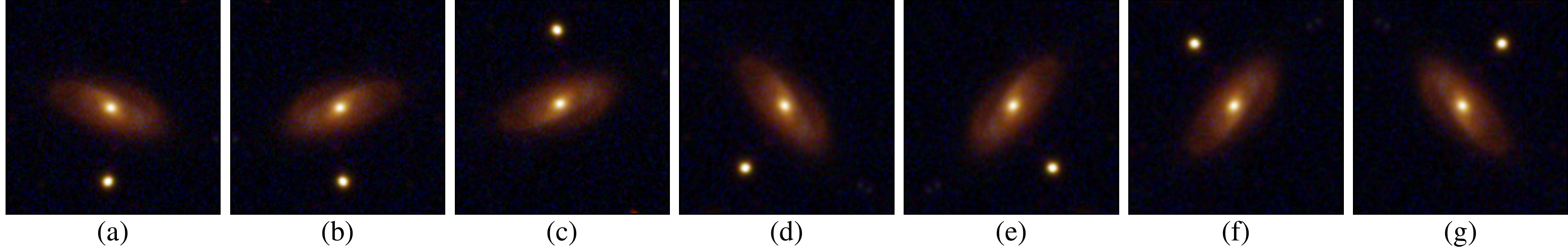}
  \caption{Exemplo do aumento artificial de dados em uma imagem original, mostrada no painel (a). Os painéis (b), (c), (d), (e), (f) e (g) contêm os resultados da equação \eqref{eq:final-transformation} substituindo $M$ por diferentes combinações das transformações da equação \eqref{eq:transformations}. Em (b) $M = V$, em (c) $M = H$, em (d) $M = R(30\degree)$, em (e) $M = V R(30\degree)$, em (f) $M = H R(30\degree)$ e em (g) $M = H V R(30\degree)$.}
  \label{fig:dataaug}
\end{figure*}

Nesta seção, explicamos sobre a preparação dos dados e como fizemos o aumento artificial dos dados para obter melhores resultados na avaliação dos modelos. Em seguida, descrevemos as redes convolucionais utilizadas: VGG, Inception Resnet, EfficientNet e DenseNet. Introduzimos o conceito de (\emph{Ensemble}) e descrevemos as técnicas usadas anteriormente e que fundamentaram nossas escolhas. Em seguida, apresentamos as principais definições das redes e parâmetros utilizados neste trabalho e por fim detalhamos como foram feitas as modelagens e treinamentos dos classificadores e do nosso meta-modelo.

\subsection{Preparação dos dados}
\label{section:preparacao}

O pré-processamento é a preparação das imagens para serem usadas pelo modelo, ou seja, é a transformação dos dados não processados em dados prontos para entrada na rede. Isso envolve representar as imagens por matrizes multidimensionais, onde cada elemento da matriz representa um pixel da imagem, e aplicar algumas transformações, especificadas a seguir.

\subsubsection{Agrupamento das bandas para confecção das imagens RGB}

Como as imagens do S-PLUS foram obtidas em 12 bandas fotométricas (listadas em \cite{oliveira2019}), para representá-las no sistema de cor RGB fizemos o seguinte mapeamento: em R colocamos as 4 bandas vermelhas r\_SDSS, i\_SDSS, J0861 e z, em G as bandas g\_SDSS J0515 e J0660 e em B as cinco bandas mais azuis u\_JAVA, J0373, J0395, J0410 e J0430 (as características destes filtros são dadas na Tabela 1 de \cite{oliveira2019}). Na combinação de bandas em cada canal, foi feita uma soma simples dos valores dos píxeis. Depois de reduzidas a três bandas, as imagens são usadas como entrada do programa Trilogy\cite{coe2012clash}\footnote{\url{https://www.stsci.edu/~dcoe/trilogy/Intro.html}}.

\subsubsection{ImageNet}

Como já mostrado em trabalhos anteriores, a inicialização dos pesos provenientes de uma rede pré-treinada usando a base de dados \emph{ImageNet}\footnote{\url{http://www.image-net.org/}} traz uma grande melhoria na precisão dos resultados da classificação. Essa base de dados possui milhões de imagens de objetos do cotidiano e já foi utilizada especificamente para a classificação de objetos astronômicos (veja por exemplo \cite{bom2021}), com excelentes resultados.

O uso deste dataset para pré-treinamento respeitou o pré-processamento utilizado originalmente pelos autores de cada rede, este procedimento este foi crucial para garantir um fit competitivo, isto é, no \textit{benchmarking} original destas redes para os dados da \emph{ImageNet}. Para a rede VGG16 (Seção \ref{section:vgg}), a ordem das bandas foi trocada de RGB para BGR, e cada banda foi centrada em zero em relação à \emph{ImageNet}, sem escalonamento, ou seja, os píxeis de cada banda tiveram o valor da média da respectiva banda \emph{ImageNet} subraído. Para a rede InceptionResNetV2 (Seção \ref{section:inceptionresnetv2}), os píxeis de entrada foram escalonados entre -1 e 1 em relação a amostra de treino. Para a rede EfficientNet (Seção \ref{section:efficientnet}), os píxeis de entrada foram escalonados entre 0 e 1 em relação à amostra de treino. E, para a rede DenseNet (Seção \ref{section:densenet}), os píxeis de entrada foram escalonados entre 0 e 1 e cada banda foi padronizada em relação à \emph{ImageNet}, isto é, os píxeis de cada banda tiveram o valor da média subtraído e o resultado foi dividido pelo desvio padrão da distribuição da respectiva banda da \emph{ImageNet}.

\begin{figure*}[!ht]
  \centering
  \includegraphics[width=\textwidth]{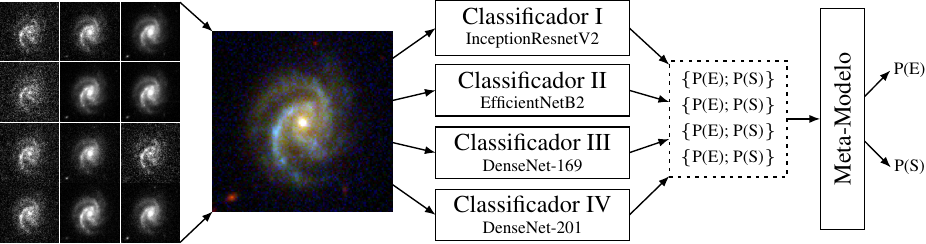}
  \caption{Diagrama que descreve a arquitetura da rede. Da esquerda para a direita, as 12 imagens de cada banda são agrupadas em uma única imagem RGB (Seção \ref{section:preparacao}), que é a entrada dos classificadores indidviduais. Estes classificadores (Seção \ref{section:classificador}) têm a função de extrair características visuais da imagem e retornar a probabilidade de ser elíptica ou espiral. O meta-modelo (Seção \ref{section:meta-modelo}) tem a função de combinar as predições dos classificadores em uma única predição final mais robusta.}
  \label{fig:arch}
\end{figure*}

\subsection{Aumento artificial de dados}

Aumento artificial de dados \cite{Larry1996} é a aplicação de transformações afins nas imagens do conjunto de treinamento, por exemplo rotação, reflexão, translação e mudança de escala. As matrizes da equação \eqref{eq:transformations} definem as transformações usadas.

\begin{equation} \label{eq:transformations}
  \begin{gathered}
    R(\theta) =
    \begin{bmatrix}
      \cos(\theta) & -\sin(\theta) & 0 \\
      \sin(\theta) & \cos(\theta)  & 0 \\
      0            & 0             & 1
    \end{bmatrix}
    \\[1.5ex]
    H =
    \begin{bmatrix}
      1 & 0  & 0 \\
      0 & -1 & 0 \\
      0 & 0  & 1
    \end{bmatrix}
    \quad
    V =
    \begin{bmatrix}
      -1 & 0 & 0 \\
      0  & 1 & 0 \\
      0  & 0 & 1
    \end{bmatrix}
  \end{gathered}
\end{equation}
onde $R(\theta)$ é a transformação rotação por um ângulo $\theta$,
$H$ é a transformação reflexão horizontal e $V$ é a transformação reflexão vertical.

Seja $M$ a matriz das transformações combinadas, $(x, y)$ a coordenada do píxel da imagem original e $(x^*, y^*)$ a coordenada transformada do píxel, as transformações nas imagens são feitas remapeando as coordenadas dos píxeis originais aplicando uma combinação das matrizes da equação \eqref{eq:transformations} em cada píxel da imagem original usando a equação \eqref{eq:final-transformation}, onde $(t_x, t_y)$ é a coordenada do centro da imagem e as matrizes ao redor de $M$ são as matrizes translação. Isso é feito para que a transformação $M$ tenha o centro da imagem como ponto de simetria.

\begin{align} \label{eq:final-transformation}
  \begin{bmatrix}
    x^* \\
    y^* \\
    1
  \end{bmatrix}
  =
  \begin{bmatrix}
    1 & 0 & t_x \\
    0 & 1 & t_y \\
    0 & 0 & 1
  \end{bmatrix}
  \ M\
  \begin{bmatrix}
    1 & 0 & -t_x \\
    0 & 1 & -t_y \\
    0 & 0 & 1
  \end{bmatrix}
  \begin{bmatrix}
    x \\
    y \\
    1
  \end{bmatrix}
\end{align}

Além disso, ainda é aplicada uma interpolação bilinear como \emph{anti-aliasing} \cite{aliasing, bilinear}. Durante o treinamento da rede, novas imagens de entrada são geradas a cada época a partir da transformação das imagens originais. A Figura \ref{fig:dataaug} mostra a imagem original, no painel (a), e diversas transformações, nos demais painéis, aplicadas substituindo $M$ da equação \eqref{eq:final-transformation} por combinações (multiplicação matricial) das transformações da equação \eqref{eq:transformations}. Tais transformações não mudam a interpretação da classe da imagem original, pois o espaço visual é invariante a elas. Logo, o objetivo de aplicar estas transformações nas imagens de entrada da rede é deixar que o algorítmo infira tal invariância, criando, assim, uma ``noção'' do espaço visual, o que resulta no aumento do potencial de generalização da rede \cite{Simard2003, CholletBook}. Frequentemente são relatados bons resultados com o uso desta técnica \cite{EfficientNetEx01, EfficientNetEx02, CNNEx04}, principalmente quando existe grande similaridade entre as classes.

\subsection{VGG}
\label{section:vgg}
A arquitetura VGG \cite{VGG16} foi criada durante a competição de classificação de imagens \textit{Large Scale Visual Recognition Challenge} \cite{ILSVRC15}. Ela se destaca por estar entre as primeiras redes a adotar, com sucesso, o escalamento em profundidade (quantidade de camadas) para aumentar o desempenho na classificação de imagens usando redes convolucionais. Ela já foi usada em diversas tarefas de classificação, como a classificação de software malicioso \cite{VGG16Ex01}, de plantas \cite{VGG16Ex02} e de tumores cerebrais \cite{VGG16Ex03}.

\subsection{InceptionResNetV2}
\label{section:inceptionresnetv2}
A arquitetura InceptionResNetV2 \cite{InceptionResNetv2} usa os blocos Inception, que são convoluções fatorizadas, introduzidos em \cite{Inception}, motivada pela construção de redes mais profundas com um menor custo computacional e menor overfitting, com a adição de conexões residuais \cite{ResNet} motivada pelo problema de dissipação do Gradiente (\textit{vanishing gradients}). Isso permite treinar redes profundas com maior acurácia e mais rápido. Esta arquitetura já foi usada, por exemplo, para classificação de imagens de satélite \cite{InceptionResNetV2Ex01}, de ultrasonografia \cite{InceptionResNetV2Ex02} e de células cancerígenas \cite{InceptionResNetV2Ex03}.

\subsection{EfficientNet}
\label{section:efficientnet}
A arquitetura EfficientNet \cite{EfficientNet} foi desenvolvida como uma resposta à questão de como escalar modelos de convolução. Foram considerados três diferentes aspectos: profundidade, largura e resolução da imagem de entrada. Em vez de dimensionar cada aspecto manualmente, o modelo implementa um escalonamento composto que equilibra os aspectos para obter melhor desempenho, com isso a rede consegue uma alta acurácia usando muito menos parâmtros e operações de ponto flutuante por segundo (\emph{FLOPS}). Esta rede já foi usada na classificação de doenças em vegetais \cite{EfficientNetEx03}, eletrocardiogramas \cite{EfficientNetEx01} e cristalização de proteínas \cite{EfficientNetEx02}.

\subsection{DenseNet}
\label{section:densenet}
A rede DenseNet \cite{DenseNet} também usa conexões residuais que conectam cada camada a todas as outras camadas seguintes, o que reduz ainda mais o número de parâmetros na rede sem perda significativa da precisão. O uso desta rede incluem predição do mapa de contato de proteínas \cite{DenseNetEx02}, classificação de músicas \cite{DenseNetEx05}, câncer de mama \cite{DenseNetEx01} e esclerose múltipla \cite{DenseNetEx03}.

\subsection{Ensemble}
Abaixo mostraremos como podemos combinar os resultados das redes descritas na última seção, usando um método de  \emph{Ensemble}, que combina os vários modelos treinados para resolver um mesmo problema. Essse ensemble foi feito inspirado em \cite{Frayman2002}, que mostraram uma grande melhoria nos resultados quando várias redes são combinadas numa regressão logística. O uso de Ensemble tem se mostrado uma importante técnica no desenvolvimento de modelos de \emph{machine learning} ainda nos anos 90, quando Hansen \& Salamon \cite{Hansen1990} mostraram que as predições feitas pela combinação de um conjunto de classificadores são frequentemente mais precisas do que as feitas somente pelo melhor classificador. Existem vários relatos de sucesso do uso desta técnica em \emph{deep learning}, dentre eles a classificação de retina humana \cite{EnsembleEx01}, melanoma \cite{EnsembleEx02} e de anomalias na Via Láctea \cite{EnsembleEx03}. Existem várias técnicas de agrupamento (\emph{Ensemble}), como \emph{Boosting} \cite{Kearns1989, Schapire1990}, \emph{Bagging} \cite{Breiman1996} e \emph{Stacking} \cite{Wolpert1992, Breiman1996b, Smyth1999}.
Este último foi o escolhido para ser usado neste trabalho. Ele se diferencia dos demais pela presença de um meta-modelo, que recebe as predições dos classificadores -- treinados individualmente -- e retorna uma predição final, como é mostrado na Figura \ref{fig:arch}.
A Figura \ref{fig:arch} mostra a arquitetura do \emph{Ensemble} a partir do processo de inferência de uma imagem. Pelo diagrama, é possível notar que o modelo é composto de duas camadas de redes neurais artificiais, a primeira é composta pelos classificadores e a segunda é composta pelo meta-modelo. As Seções \ref{section:classificador} e \ref{section:meta-modelo} detalham o desenvolvimento da primeira e segunda camadas respectivamente.

Uma outra etapa importante no desenvolvimento de redes neurais artificiais é o ajuste dos hiper-parâmetros, alguns deles mostrados na Seção \ref{section:hyperparam}.

\subsection{Definições das redes e parâmetros utilizados}
\label{section:hyperparam}

Nesta seção descrevemos as definições dos principais conceitos, no contexto de deep learning, que serão úteis para o entendimento dos métodos aqui utilizados. A função de ativação, função de custo, o otimizador, o learning rate, o número de épocas, além do número de camadas dos modelos, são importantes parâmetros responsáveis pela contrução do modelo definido a seguir.

\begin{description}
  \item[Função de ativação] \hfill \\
        A função de ativação é responsável por adicionar não-linearidade à rede. Sem ela, a saída de uma camada seria apenas uma transformação linear dos dados de entrada e a rede não seria beneficiada pelo empilhamento de diversas camadas lineares, pois isso não aumentaria o espaço de hipóteses. Logo, a função de ativação viabiliza representações mais complexas da rede, uma vez que define a complexidade de um modelo e, consequentemente, sua capacidade de generalização \cite{CholletBook}. Neste trabalho, a função $\rm{ReLU}(x) := \rm{max}(0, x)$ é usada nas camadas densas dos classificadores, a equação tangente hiperbólica é usada nas camadas densas do meta-modelo e a função Softmax \cite{Bridle1990} foi usada na última camada, tanto dos classificadores quanto do meta-modelo.

  \item[Função de Custo] \hfill \\
        A função de custo é utilizada com o objetivo de determinar o quão longe o modelo está do esperado, definindo a necessidade de atualização dos pesos da rede. Utilizamos a função Entropia Cruzada (\emph{Cross-Entropy})
        \begin{equation}
          \label{eq:custo}
          J = \sum_{i=1}^{C} y_i \cdot \rm{log}(\hat{y}_i)
        \end{equation}
        onde $y_i$ representa a probabilidade da classe dada pelo conjunto de treinamento do objeto $i$ e $\hat{y}_i$ representa a previsão da rede para este mesmo objeto.

  \item[Otimizador] \hfill \\
        O otimizador é um algorítmo iterativo com objetivo de minimizar a função de custo. Uma escolha típica é o método de gradiente descente e suas demais variações. Este tipo de algoritmo tem um parâmetro livre relacionado ao passo da iteração conhecido como taxa de aprendizado ou \textit{learning rate}. Neste trabalho foram testados diversos algoritmos considerados como estado-da-arte dos otimizadores como Adam \cite{Adam}, NAdam \cite{NAdam}, RAdam \cite{RAdam} e RMSprop \cite{RMSprop}.

  \item[Número de Épocas] \hfill \\
        O Número de épocas se referem a quantidade de vezes que o dataset de treino foi utilizado completamente no processo de otimização iterativa da função de custo. Um número de épocas adequado é necessário para que a função de custo seja minimizada.

  \item[Tamanho do Batch] \hfill \\
        O processo de otimização acontece em batches, cada iteração para minimizar a função custo é realizada com um número fixo de amostras, quando todas as amostras de treino são utilizadas se completa uma época.

  \item[Unidades de neurônios na última camada]
        \hfill \\
        A última camada da rede antes da camada de saída é responsável por condensar toda a informação extraída da rede para o processo de classificação final. Por esta razão, a quantidade de neurônios nessa camada pode ser particularmente sensível para a performance da rede. Neste trabalho utilizamos diferentes valores de neurônios para encontrar a quantidade que pode gerar a melhor performance.

  \item[Dropout]
        \hfill \\
        \emph{Dropout} \cite{dropout} é uma técnica de regularização muito utilizada em redes neurais por seu bom desempenho e baixo custo computacional. Aplicar esta regularização em uma camada consiste em eliminar aleatoriamente uma taxa dos neurônios de saída desta camada durante o treinamento, sendo geralmente escolhido um valor entre 0.2 e 0.5 para esta taxa \cite{CholletBook}.
\end{description}

\subsection{Modelagem e Treinamento dos Classificadores}
\label{section:classificador}

\begin{table}[!ht]
  \centering
  \caption{Variação dos parâmetros dos classificadores com arquitetura InceptionResNetV2 usando \emph{oversample}.}
  \label{tab:First_step_hip}
  \begin{tabular}{cccc}
    \toprule
    Parâmetro                          & Valor             & ROC AUC       & PR AUC        \\
    \midrule
    \multirow{4}{*}{Otimizador}        & Adam              & 0.983551      & 0.984164      \\
                                       & RAdam             & 0.981333      & 0.982089      \\
                                       & NAdam             & \bf{0.985507} & \bf{0.986090} \\
                                       & RMSprop           & 0.976397      & 0.977327      \\
    \midrule
    \multirow{7}{*}{\shortstack{\emph{Learning}                                            \\\emph{Rate}}} & $5 \cdot 10^{-6}$ & 0.970411 & 0.971590 \\
                                       & $1 \cdot 10^{-5}$ & 0.978090      & 0.978931      \\
                                       & $5 \cdot 10^{-5}$ & \bf{0.986820} & \bf{0.987339} \\
                                       & $1 \cdot 10^{-4}$ & 0.986380      & 0.986919      \\
                                       & $5 \cdot 10^{-4}$ & 0.985809      & 0.986337      \\
                                       & $1 \cdot 10^{-3}$ & 0.981214      & 0.979992      \\
                                       & $5 \cdot 10^{-3}$ & 0.977913      & 0.976907      \\
    \midrule
    \multirow{5}{*}{\emph{Batch Size}} & 32                & 0.974966      & 0.972293      \\
                                       & 64                & 0.981530      & 0.980156      \\
                                       & 128               & 0.976305      & 0.974936      \\
                                       & 192               & \bf{0.985100} & \bf{0.985659} \\
                                       & 256               & 0.983728      & 0.984385      \\
    \midrule
    \multirow{6}{*}{Unidades}          & 128/2             & 0.977434      & 0.978307      \\
                                       & 256/2             & 0.979423      & 0.980226      \\
                                       & 512/2             & \bf{0.986420} & \bf{0.986917} \\
                                       & 1024/2            & 0.981792      & 0.982526      \\
                                       & 256/128/2         & 0.979862      & 0.980575      \\
                                       & 1024/256/128/2    & 0.982252      & 0.982909      \\
    \bottomrule
  \end{tabular}
\end{table}

A estrutura de cada classificador é composta por um extrator de características ligado a um bloco de camadas densamente conectadas, também chamado de bloco de predição. O extrator de características é uma rede neural convolucional usada como camada de abstração das características visuais, detectando padrões como geometrias, contrastes, texturas e cores na imagem de entrada. Enquanto que o bloco de predição recebe estas característcas e retornam as probabilidades da imagem pertencer à classe elíptica e espiral.

Para o extrator de características, foram testadas arquiteturas conhecidas de redes convolucionais descritas anteriormente: InceptionResNet, EfficientNet, DenseNet e VGG. Cada uma destas arquiteturas foi pré-treinada usando a base de dados \emph{ImageNet}.

Para as camadas densas, foram testadas várias configurações de escalamento, tanto em largura (quantidade de neurônios) quanto em profundidade (quantidade de camadas). As configurações testadas consistem de $m$ camadas ocultas com $n$ unidades de neurônios ($m$ variando de 0 a 3 e $n = 2^t$, com $t$ variando de 6 a 10) ligadas a uma camada final com 2 unidades de neurônios e função de ativação \emph{Softmax}. Cada unidade desta última camada representa a probabilidade do objeto pertencer a cada classe \cite{Goodfellow2016}, i.e., elíptica e espiral. Por isso o número de neurônios é igual ao número de classes e suas configurações não foram variadas.

Além disso, as camadas ocultas foram regularizadas com \emph{dropout}, que garante uma melhora na capacidade de generalização da rede \cite{dropout2}. A avaliação dos modelos nos conjuntos de treinamento e de validação mostraram que a melhor taxa de \emph{dropout} é $0.4$, superando a avalição das redes sem regularizador e das redes com outras taxas de \emph{dropout} no intervalo entre 0.1 e 0.6. Do mesmo modo, a função de ativação \emph{ReLU} garantiu as melhores avaliações dos modelos quando comparadas com outras funções de ativação, como tanh e \emph{ELU}.

Outras configurações da rede também foram testadas, tanto relacionadas aos dados, como a amostragem, quanto relacionadas ao treinamento, como o tamanho do \emph{batch}, o algorítmo de otimização e a sua taxa de aprendizagem. Com a variação de um parâmetro por vez, é possível detectar os valores que contribuem para a melhor avaliação da rede, como visto na Tabela \ref{tab:First_step_hip}. Nesta tabela, é mostrado que um modelo treinado usando o otimizador NAdam com uma taxa de aprendizagem de $5 \cdot 10^{-5}$ e 192 exemplares por \emph{batch}, além de um bloco de predição contendo uma camada de 512 unidades de neurônios seguida de uma camada com 2 unidades, obteve a melhor avaliação no conjunto de validação em relação às outras configurações. Isso mostra como foram determinados os parâmetros do Modelo A da Tabela \ref{tab:best_hip}, mostrada na Seção \ref{section:resultados}. Os parâmetros para os demais modelos foram obtidos analogamente.

\begin{table*}[!ht]
  \caption{Avaliação, no conjunto de teste, dos classificadores com os melhores hiperparâmetros.}
  \label{tab:best_hip}
  \begin{tabular}{*{6}{c@{\hskip 9pt}}c}
    \toprule
    Parâmetro         & Modelo A          & Modelo B            & Modelo C          & Modelo D          & Modelo E          & Modelo F          \\
    \midrule
    Arquitetura       & InceptionResNetV2 & EfficientNet-B2     & DenseNet-169      & DenseNet-201      & EfficientNet-B7   & VGG16             \\
    Learning Rate     & $5\cdot10^{-5}$   & $5\cdot10^{-5}$     & $5\cdot10^{-5}$   & $5\cdot10^{-5}$   & $5\cdot10^{-5}$   & $1\cdot10^{-5}$   \\
    \emph{Batch Size} & 192               & 64                  & 192               & 192               & 192               & 192               \\
    Amostragem        & \emph{Oversample} & \emph{Class Weight} & \emph{Oversample} & \emph{Oversample} & \emph{Oversample} & \emph{Oversample} \\
    Optimizador       & NAdam             & Adam                & NAdam             & NAdam             & RAdam             & NAdam             \\
    Unidades          & 512/2             & 256/128/2           & 1024/2            & 1024/2            & 512/2             & 1024/2            \\
    \midrule
    Acurácia (\%)     & $92.75 \pm 0.77$  & $93.12 \pm 0.94$    & $93.12 \pm 0.88$  & $93.48 \pm 0.84$  & $91.30 \pm 1.06$  & $93.48 \pm 1.38$  \\
    $F_1$-Score (\%)  & $96.56 \pm 0.65$  & $96.26 \pm 1.04$    & $97.44 \pm 0.77$  & $97.74 \pm 0.91$  & $94.79 \pm 1.02$  & $96.37 \pm 0.97$  \\
    ROC AUC (\%)      & $97.63 \pm 0.15$  & $97.89 \pm 0.21$    & $97.57 \pm 0.14$  & $97.90 \pm 0.27$  & $96.84 \pm 0.19$  & $96.69 \pm 1.14$  \\
    PR AUC (\%)       & $97.73 \pm 0.14$  & $97.87 \pm 0.30$    & $97.38 \pm 0.20$  & $97.96 \pm 0.30$  & $96.94 \pm 0.17$  & $95.85 \pm 1.45$  \\
    \bottomrule
  \end{tabular}
\end{table*}

\subsection{Modelagem e Treinamento do Meta-Modelo}
\label{section:meta-modelo}

\begin{figure}[ht]
  \centering
  \includegraphics[width=\linewidth]{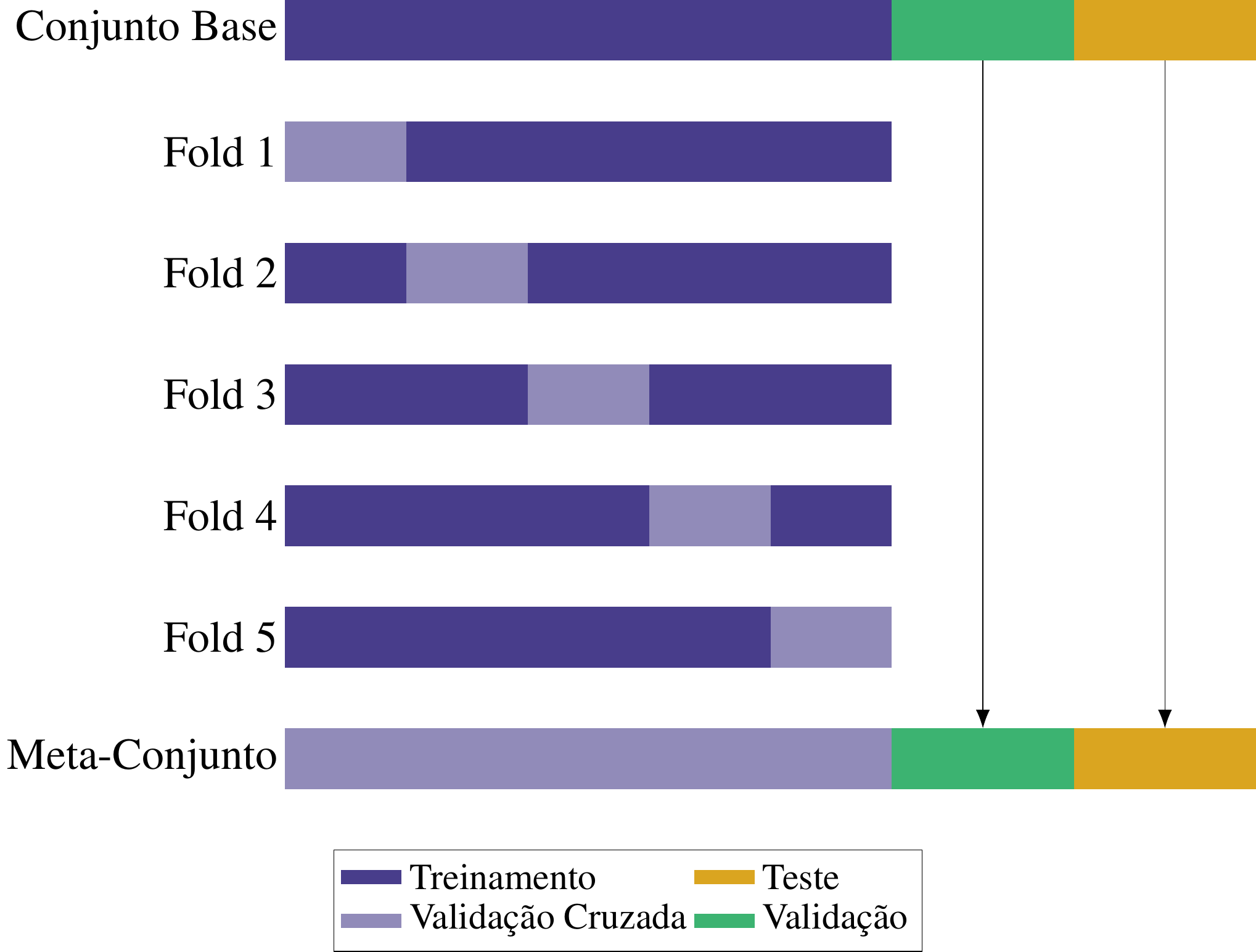}
  \caption{O diagrama representa a construção do conjunto de treinameto do meta-modelo (meta-conjunto) usado o método de validação cruzada \emph{k-fold}.}
  \label{fig:meta-kfold}
\end{figure}

O objetivo do meta-modelo é receber as predições de cada um dos classificadores e retornar uma predição final. Para isso, foi usada uma rede neural com estrutura composta por uma camada de entrada de 8 unidades de neurônios seguidas de duas camadas ocultas de 128 unidades e uma camada de saída com 2 unidades. Assim como nos classificadores, diversas configurações de hiperparâmetros foram testadas para o meta-modelo. Sendo que a configuração que gerou o melhor resultado na avaliação foi usando a função de ativação tangente hiperbólica nas camadas ocultas, \emph{batch size} de 32 e o otimizador Adam com uma taxa da aprendizagem de $1 \cdot 10^{-5}$.

Como a entrada do meta-modelo são as predições dos classificadores, o conjunto de treinamento deve ser gerado a partir destes valores. Sendo assim, para criação do conjunto de treinamento do meta-modelo, foi usado um processo iterativo chamado validação cruzada k-fold (\emph{k-fold cross-validation}). Neste método, o conjunto de treinamento é dividido em $k$ amsotras de mesmo tamanho, sendo uma delas usada obter as predições enquanto as $k-1$ restantes são usadas como treino. O processo se repete por $k$ vezes, variando a amostra usada para predição. No final da repetição deste processo para cada classificador, é obtido um conjunto de treinamento para o meta-modelo do mesmo tamanho do conjunto de treinamento original, com a vantagem que as predições foram feitas em uma amostra não usada no treinamento, como mostrado na Figura \ref{fig:meta-kfold}. Para o treinamento deste meta-modelo, foi usado um valor de $k=5$.

Para reduzir o impacto da variância das predições do conjunto de treinamento na predição final do meta-modelo, o valor usado como entrada do meta-modelo foi a mediana de 12 repetições da validação cruzada feitas para cada um dos classificadores. Além disso, os dados de entrada ainda receberam um pré-processamento, eles foram subtraídos da média e divididos pelo desvio padrão em relação às predições de cada classificador. Ao contrário dos classificadores, as camadas ocultas do meta-modelo não foram regularizadas com \emph{dropout}.

\section{Resultados}
\label{section:resultados}

\begin{figure*}[!ht]%
  \centering
  \begin{subfigure}{.5\linewidth}%
    \includegraphics[width=\linewidth]{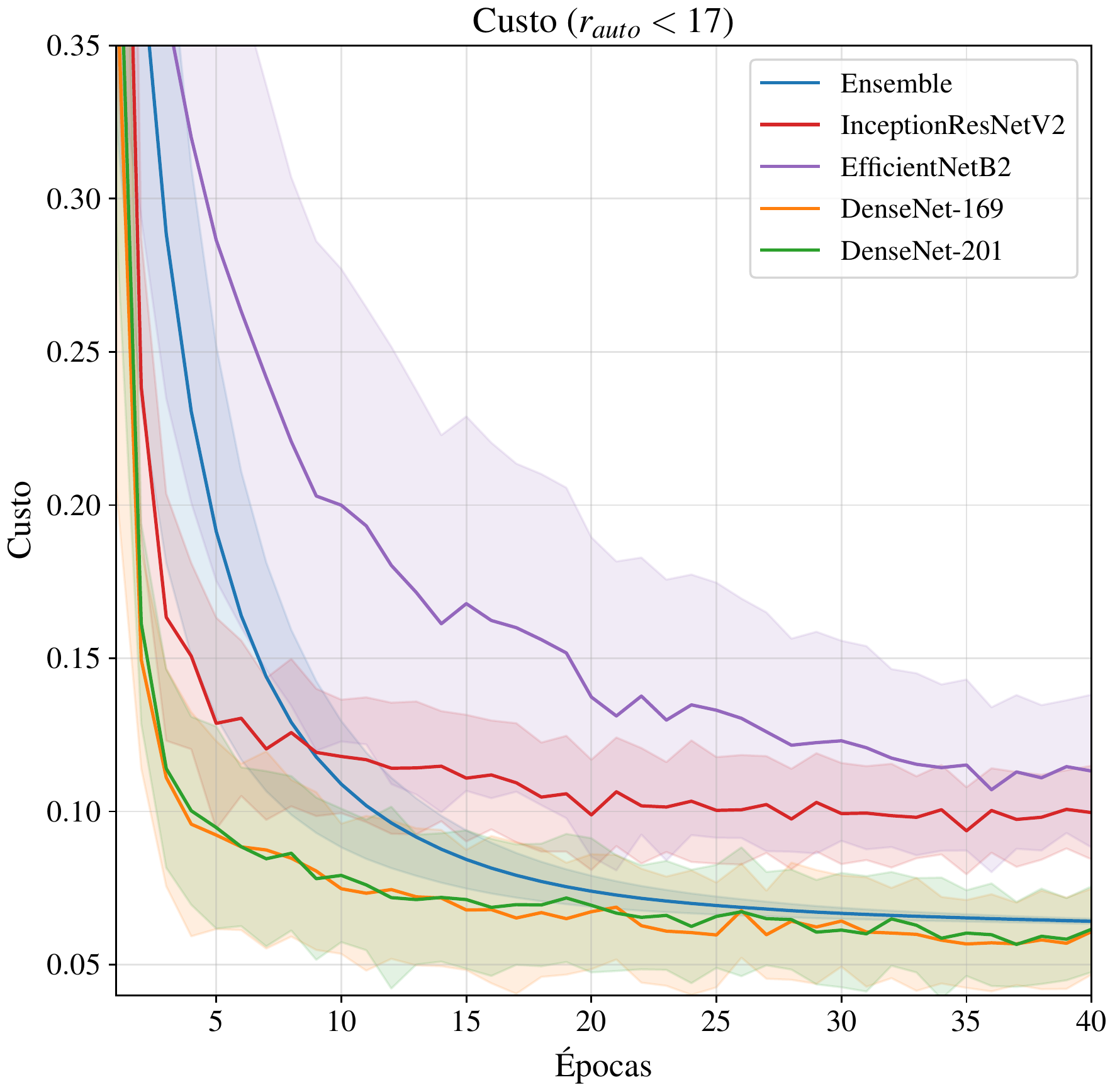}
  \end{subfigure}%
  \begin{subfigure}{.5\linewidth}%
    \includegraphics[width=\linewidth]{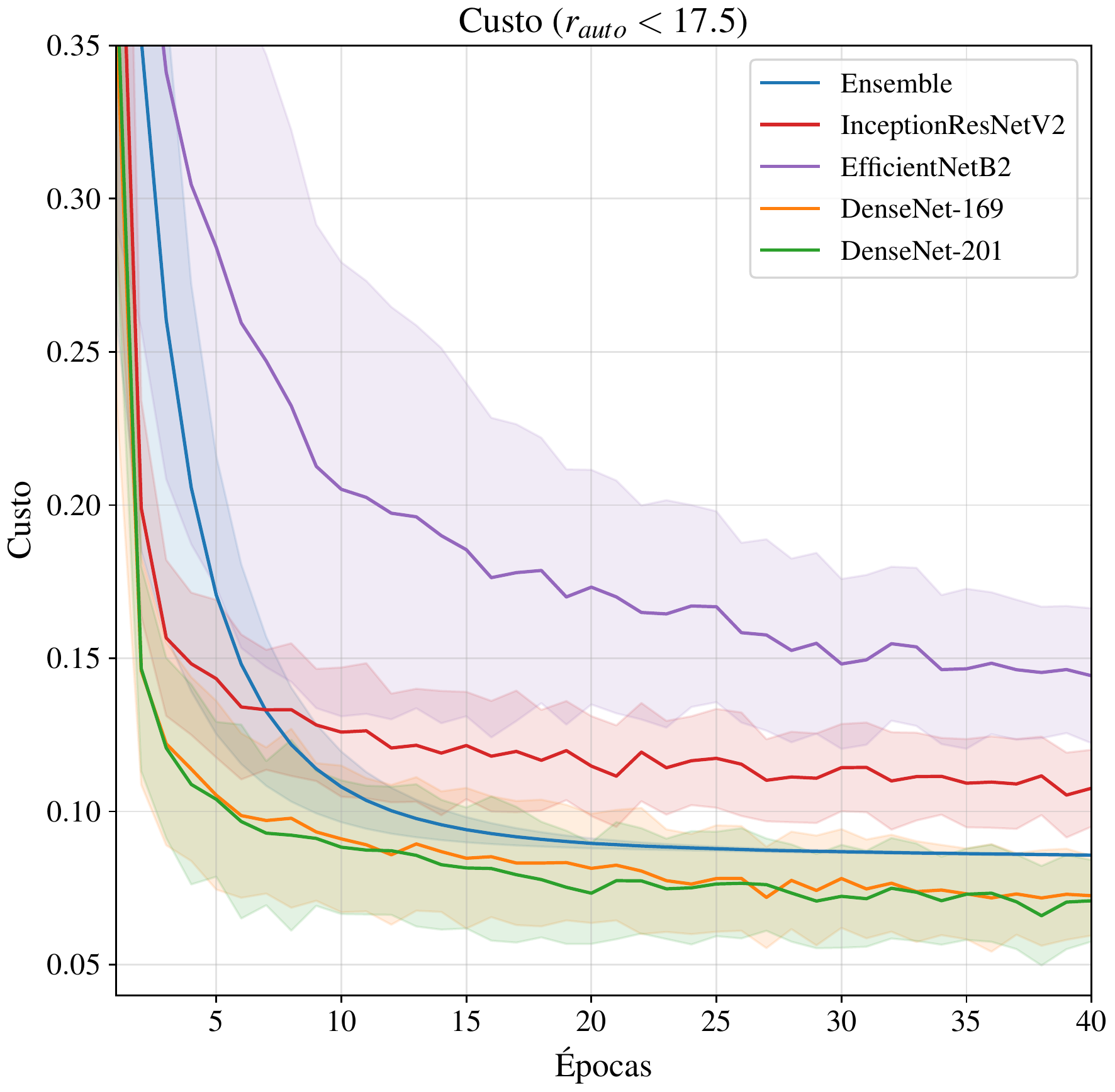}
  \end{subfigure}\\
  \begin{subfigure}{.5\linewidth}%
    \includegraphics[width=\linewidth]{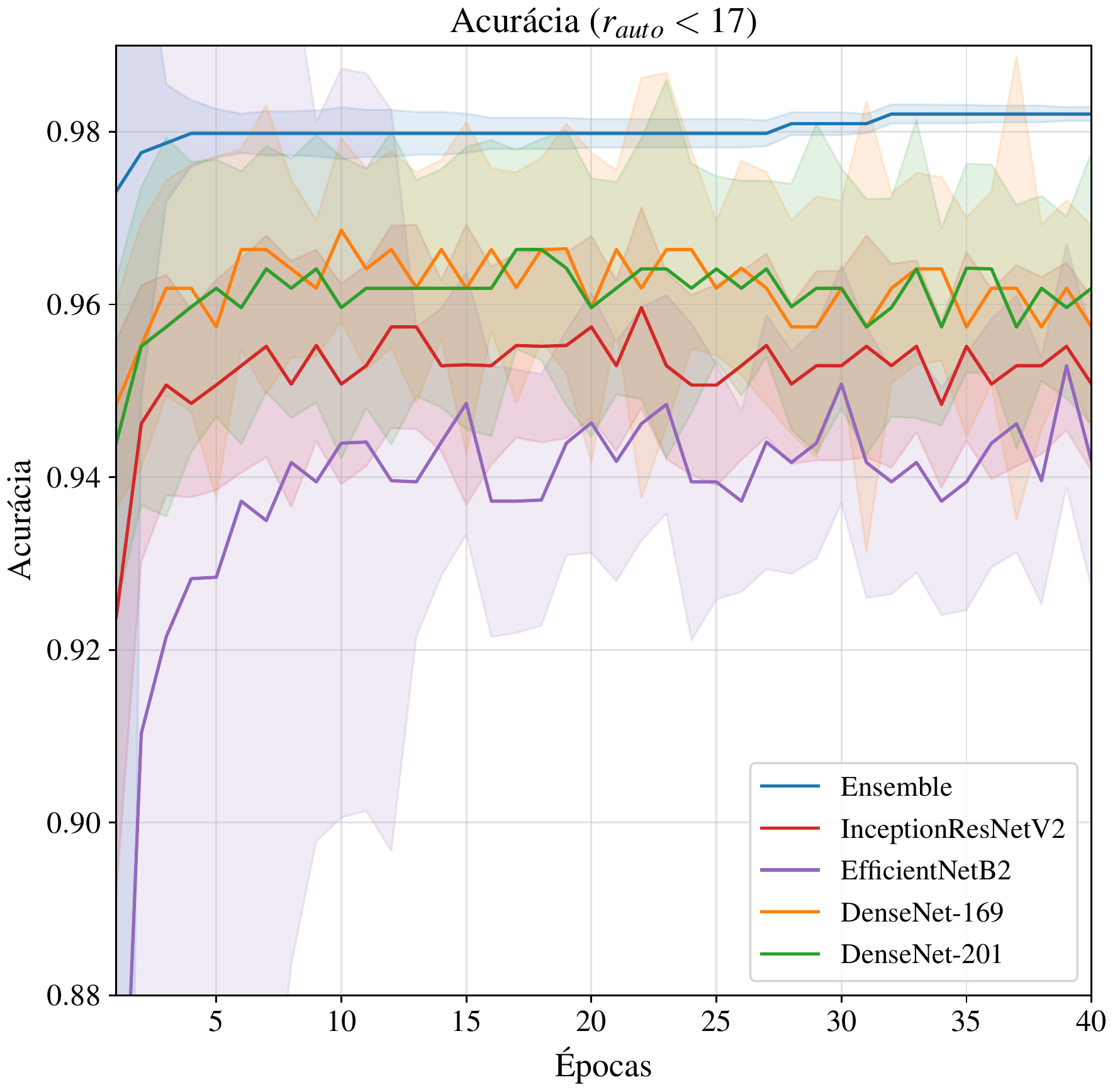}
  \end{subfigure}%
  \begin{subfigure}{.5\linewidth}%
    \includegraphics[width=\linewidth]{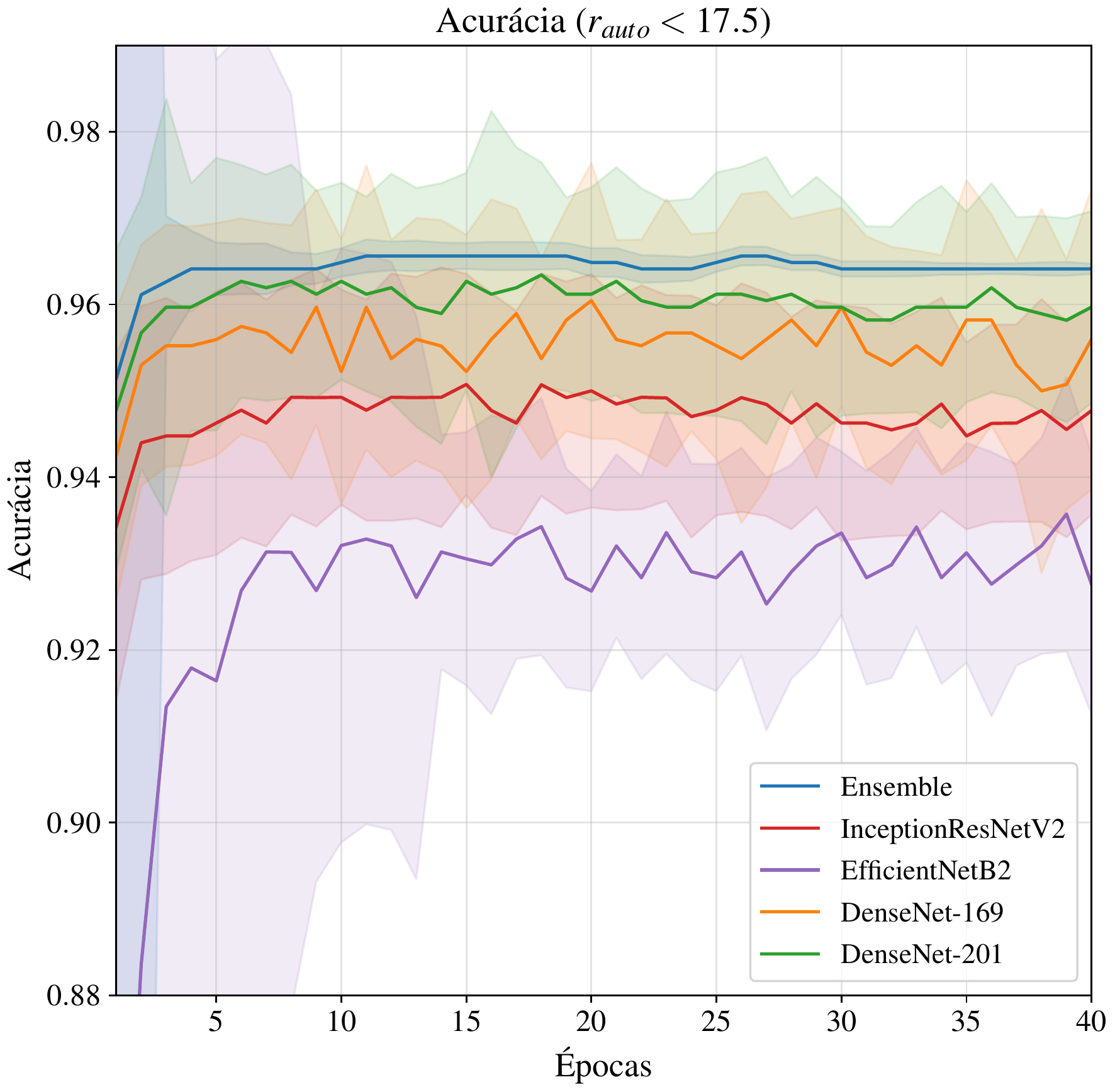}
  \end{subfigure}
  \caption{Os gráficos mostram a avaliação dos classificadores individuais e do Ensemble  para modelos treinados com $r_{auto} < 17$ (esquerda) e $r_{auto} < 17.5$ (direita). É mostrada a função custo no conjunto de treinamento (painéis de cima) e a acurácia no conjunto de validação (painéis de baixo) para cada época de treinamento. As linhas contínuas representam a mediana de 60 medições, enquanto que as regiões sombreadas representam o desvio padrão.}%
  \label{fig:train-metrics}%
\end{figure*}

\begin{figure*}[!ht]%
  \centering
  \begin{subfigure}{.5\linewidth}%
    \includegraphics[width=.98\linewidth]{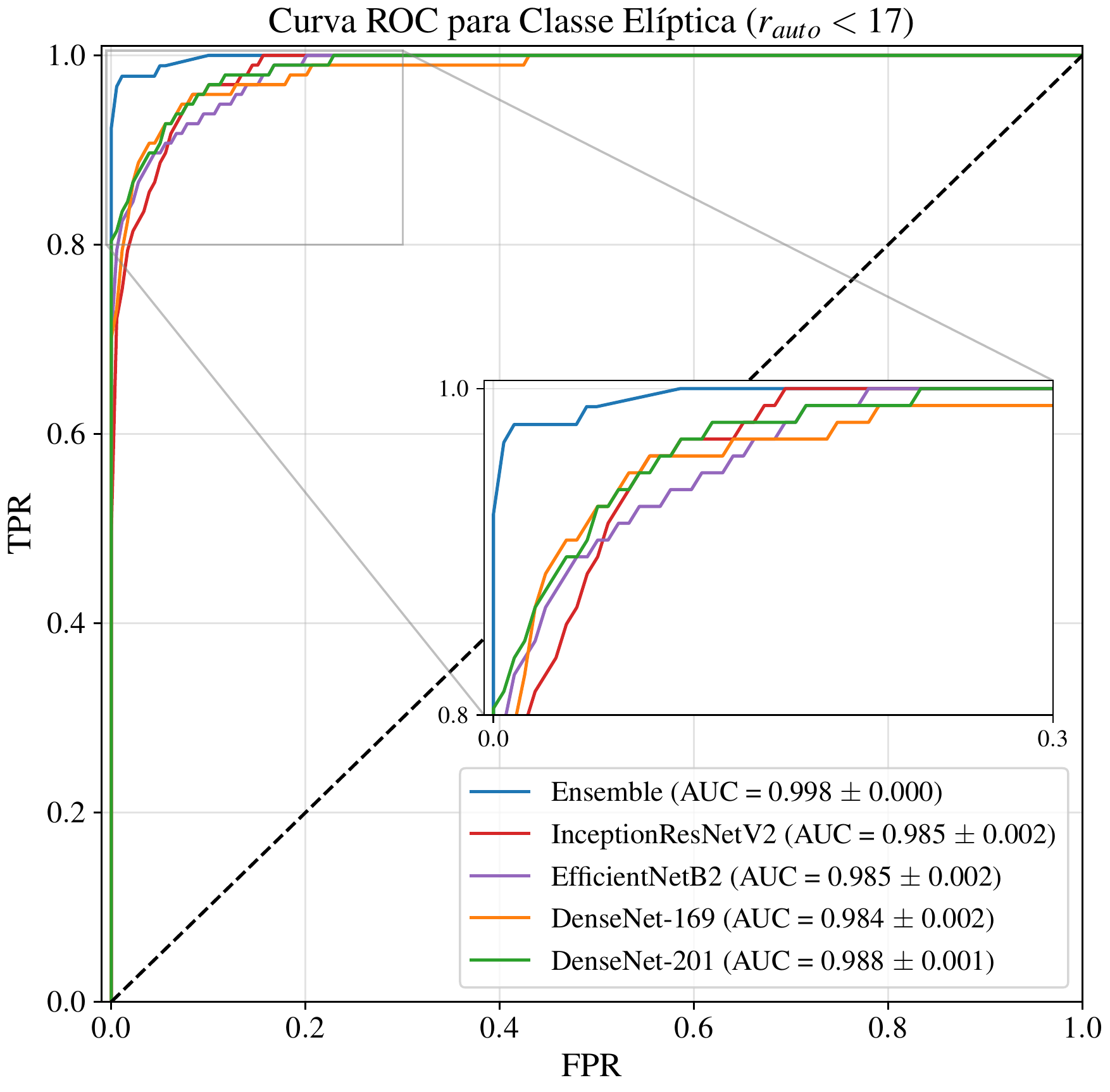}
  \end{subfigure}%
  \begin{subfigure}{.5\linewidth}%
    \includegraphics[width=.98\linewidth]{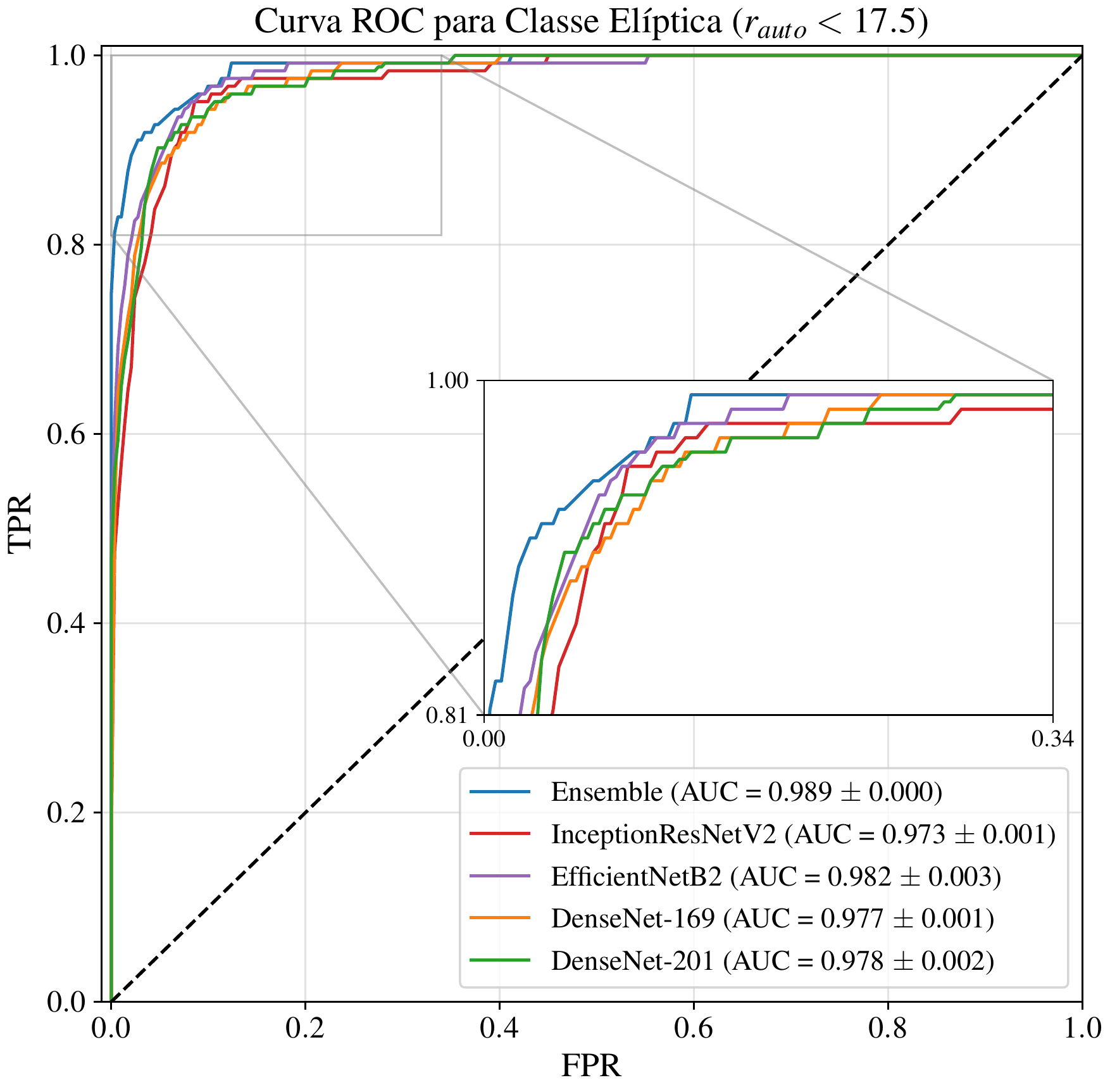}
  \end{subfigure}\\
  \begin{subfigure}{.5\linewidth}%
    \includegraphics[width=.98\linewidth]{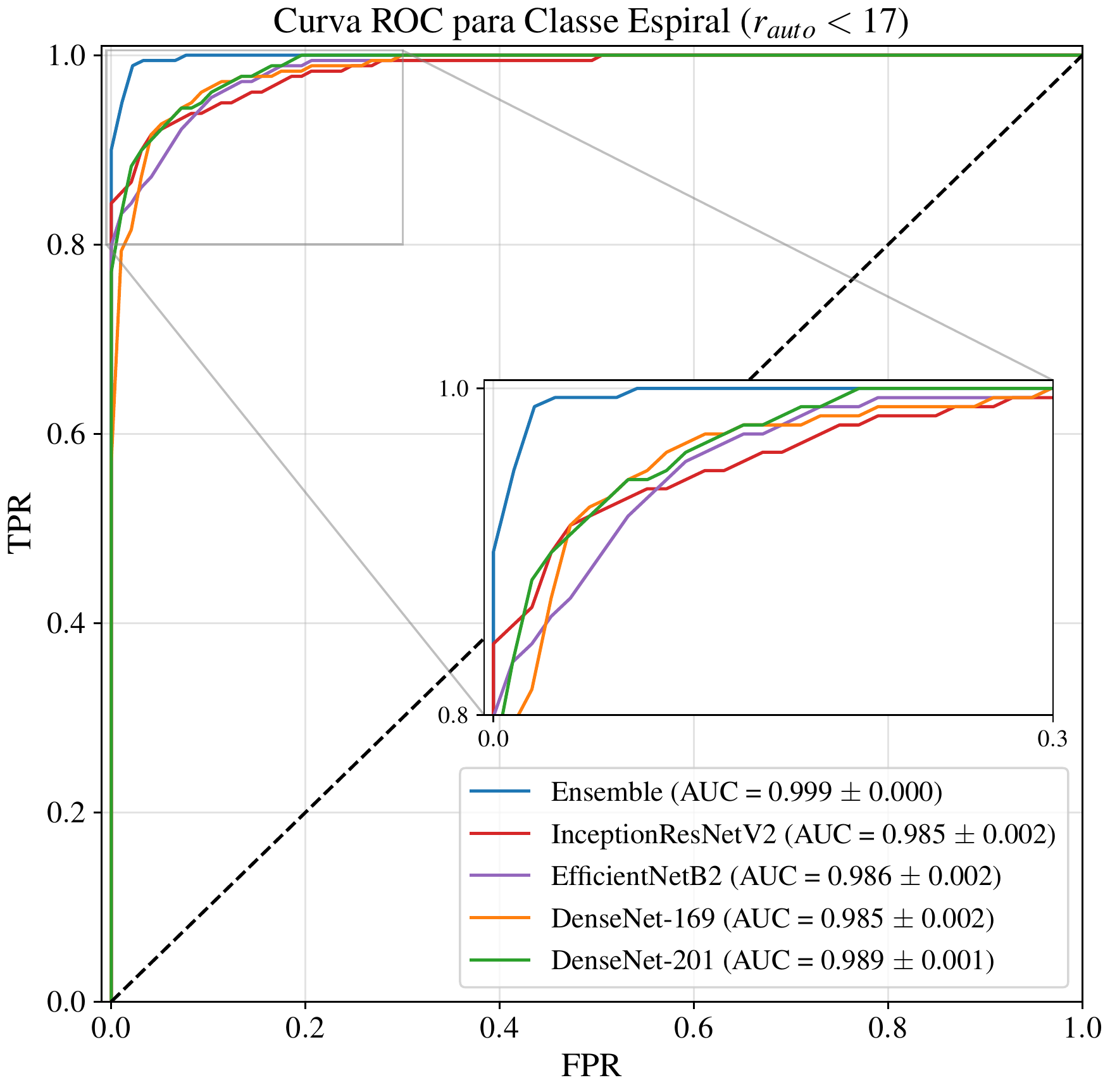}
  \end{subfigure}%
  \begin{subfigure}{.5\linewidth}%
    \includegraphics[width=.98\linewidth]{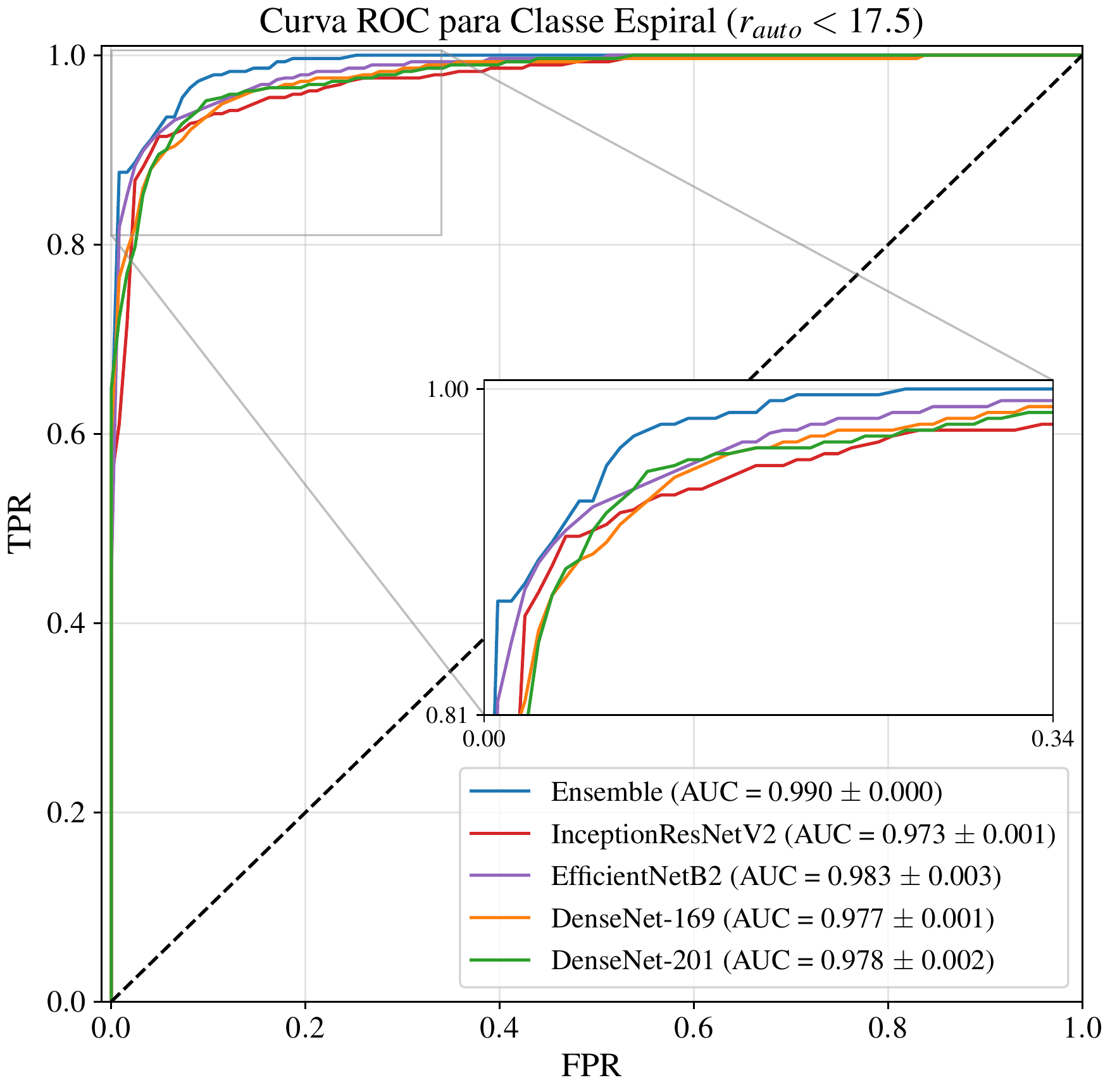}
  \end{subfigure}\\
  \caption{Curvas ROC dos classificadores individuais e do \emph{Ensemble} para os modelos treinados com $r_{auto} < 17$ (esquerda) e $r_{auto} < 17.5$ (direita) separadas pelas classes Elíptica (nos painéis de cima) e Espiral (nos painéis de baixo). As linhas contínuas mostram a mediana de 60 curvas e o valor de AUC, na legenda, representa a mediana da área abaixo destas curvas e seu respectivo desvio padrão.}%
  \label{fig:roc}
\end{figure*}

\begin{figure*}[!ht]%
  \centering
  \begin{subfigure}{.5\linewidth}%
    \includegraphics[width=.98\linewidth]{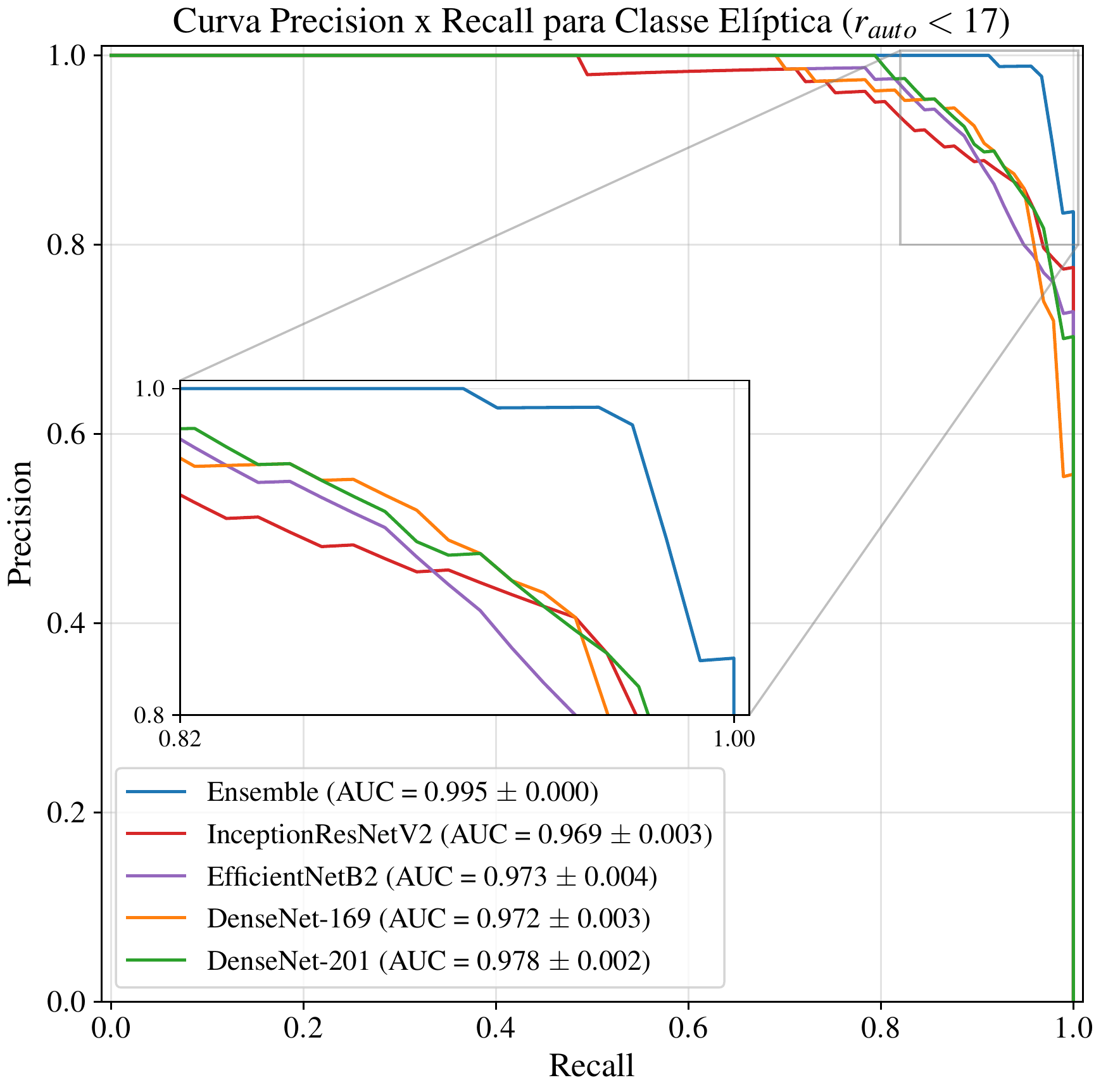}
  \end{subfigure}%
  \begin{subfigure}{.5\linewidth}%
    \includegraphics[width=.98\linewidth]{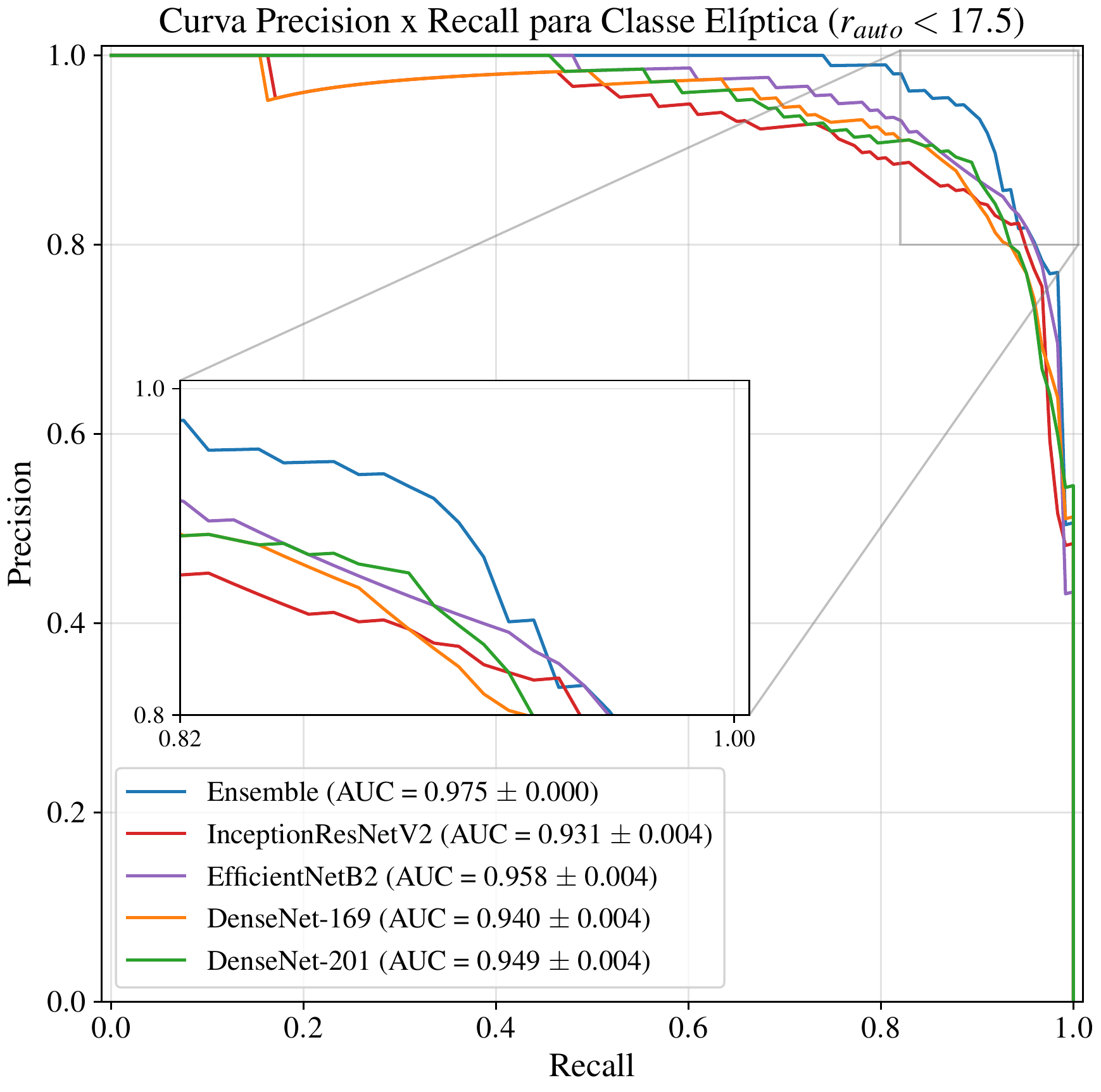}
  \end{subfigure}\\
  \begin{subfigure}{.5\linewidth}%
    \includegraphics[width=.98\linewidth]{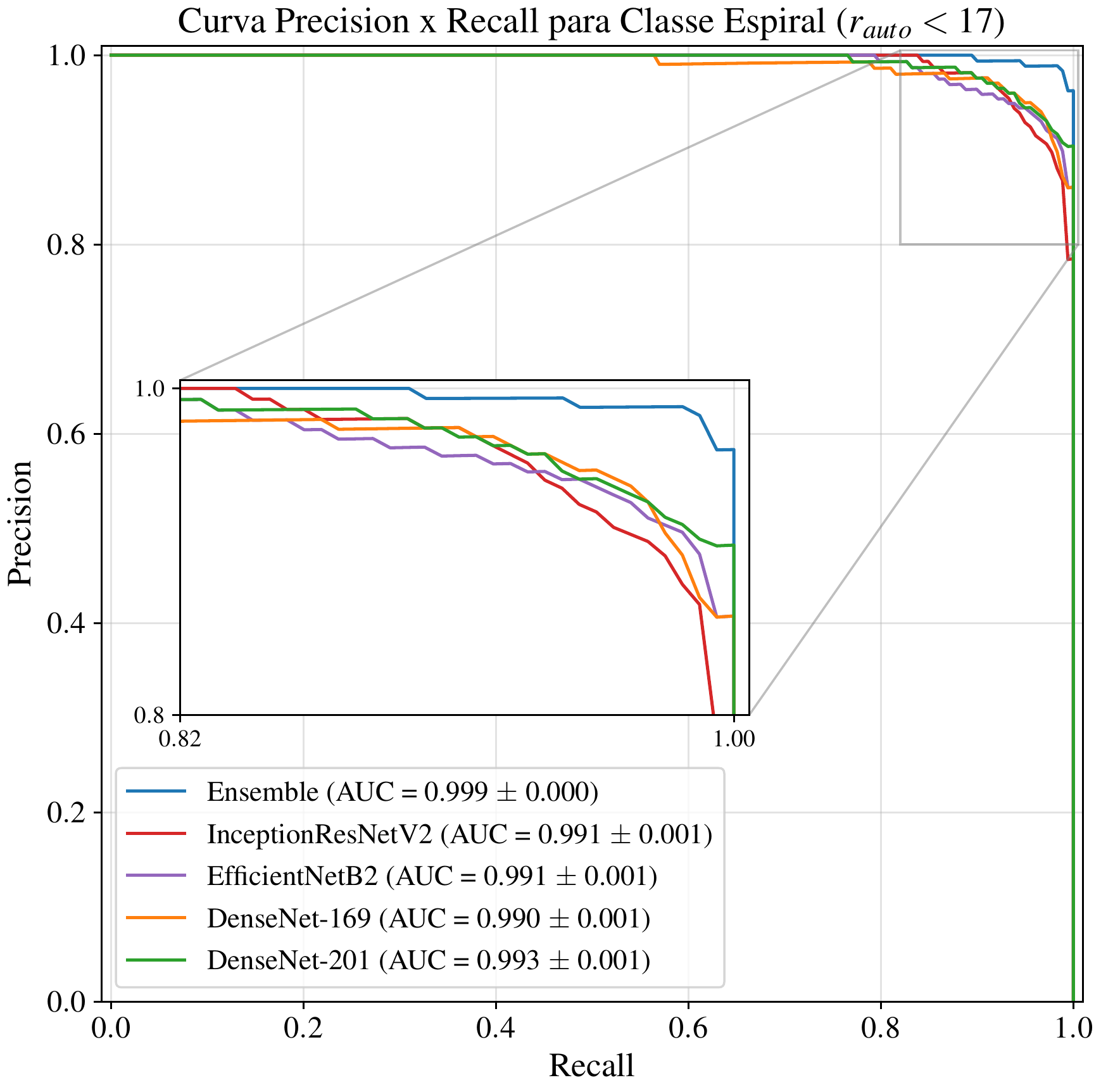}
  \end{subfigure}%
  \begin{subfigure}{.5\linewidth}%
    \includegraphics[width=.98\linewidth]{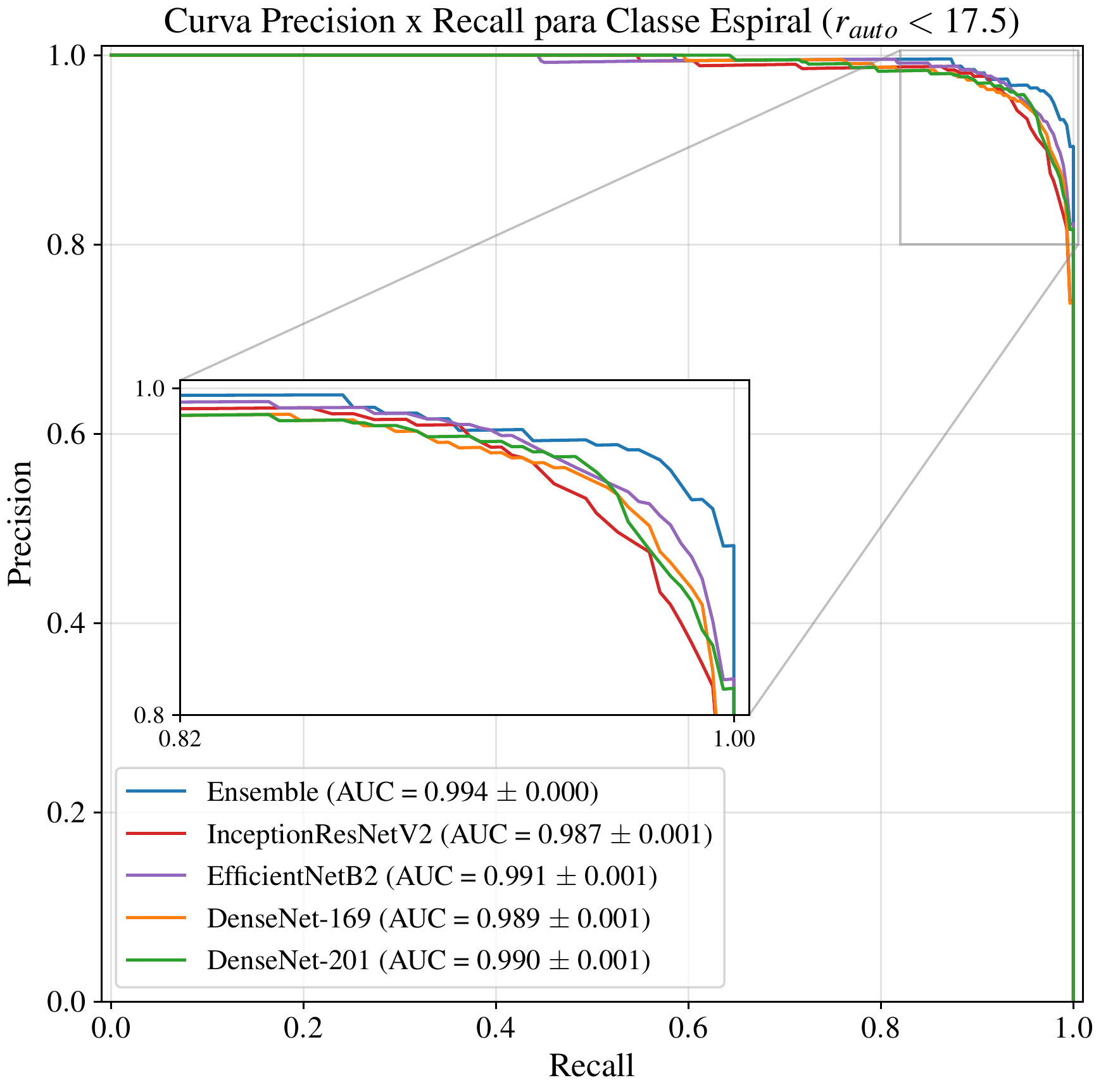}
  \end{subfigure}\\
  \caption{Curvas \emph{Precision x Recall} dos classificadores individuais e do Ensemble para os modelos treinados com $r_{auto} < 17$ (esquerda) e $r_{auto} < 17.5$ (direita) separadas pelas classes Elíptica (nos painéis de cima) e Espiral (nos painéis de baixo). As linhas contínuas mostram a mediana de 60 curvas e o valor de AUC, na legenda, representa a mediana da área abaixo destas curvas e seu respectivo desvio padrão.}%
  \label{fig:pr}
\end{figure*}

\begin{figure*}[!ht]%
  \centering
  \begin{subfigure}{.5\linewidth}
    \includegraphics[width=\linewidth]{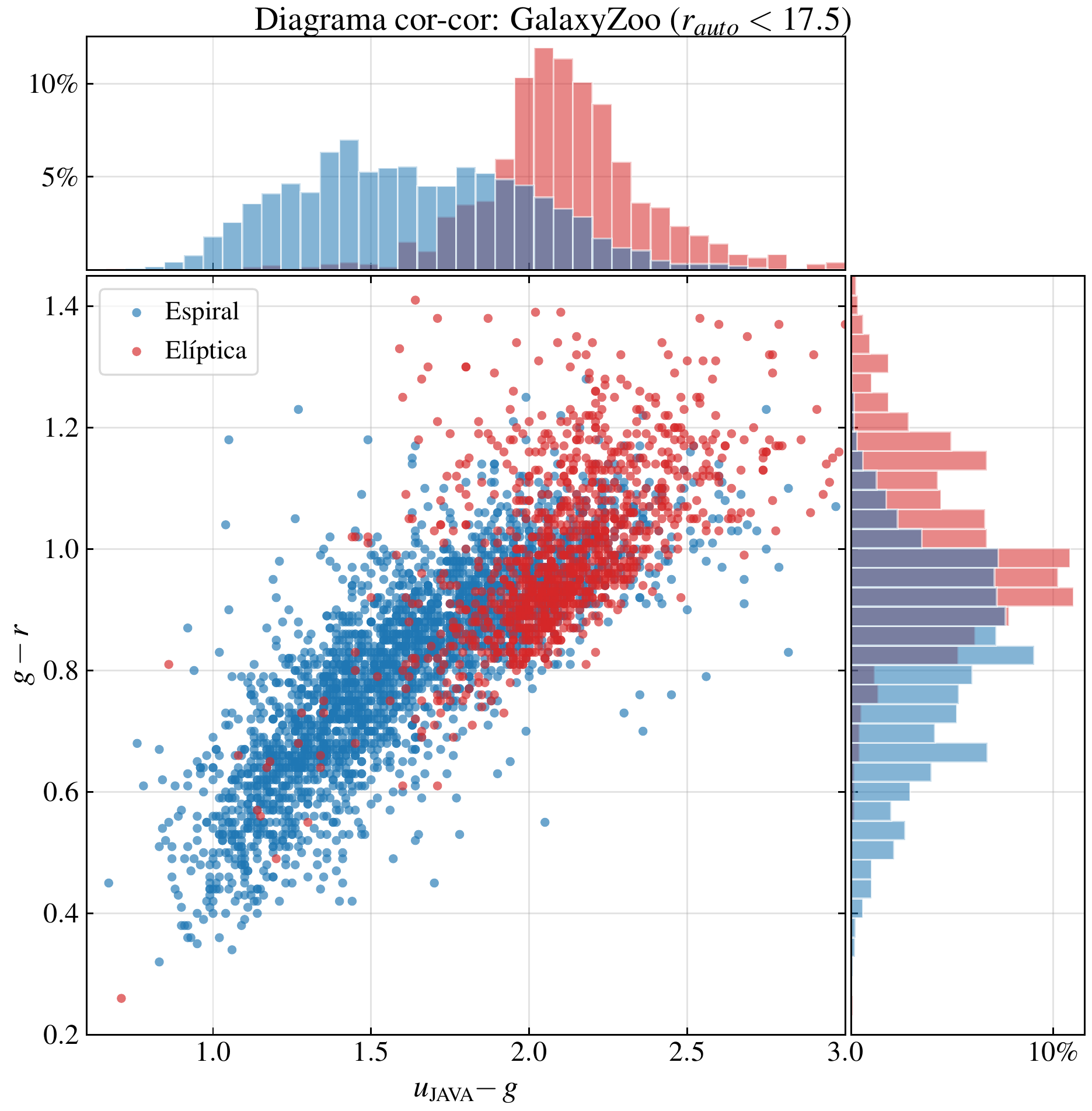}
  \end{subfigure}%
  \begin{subfigure}{.5\linewidth}
    \includegraphics[width=\linewidth]{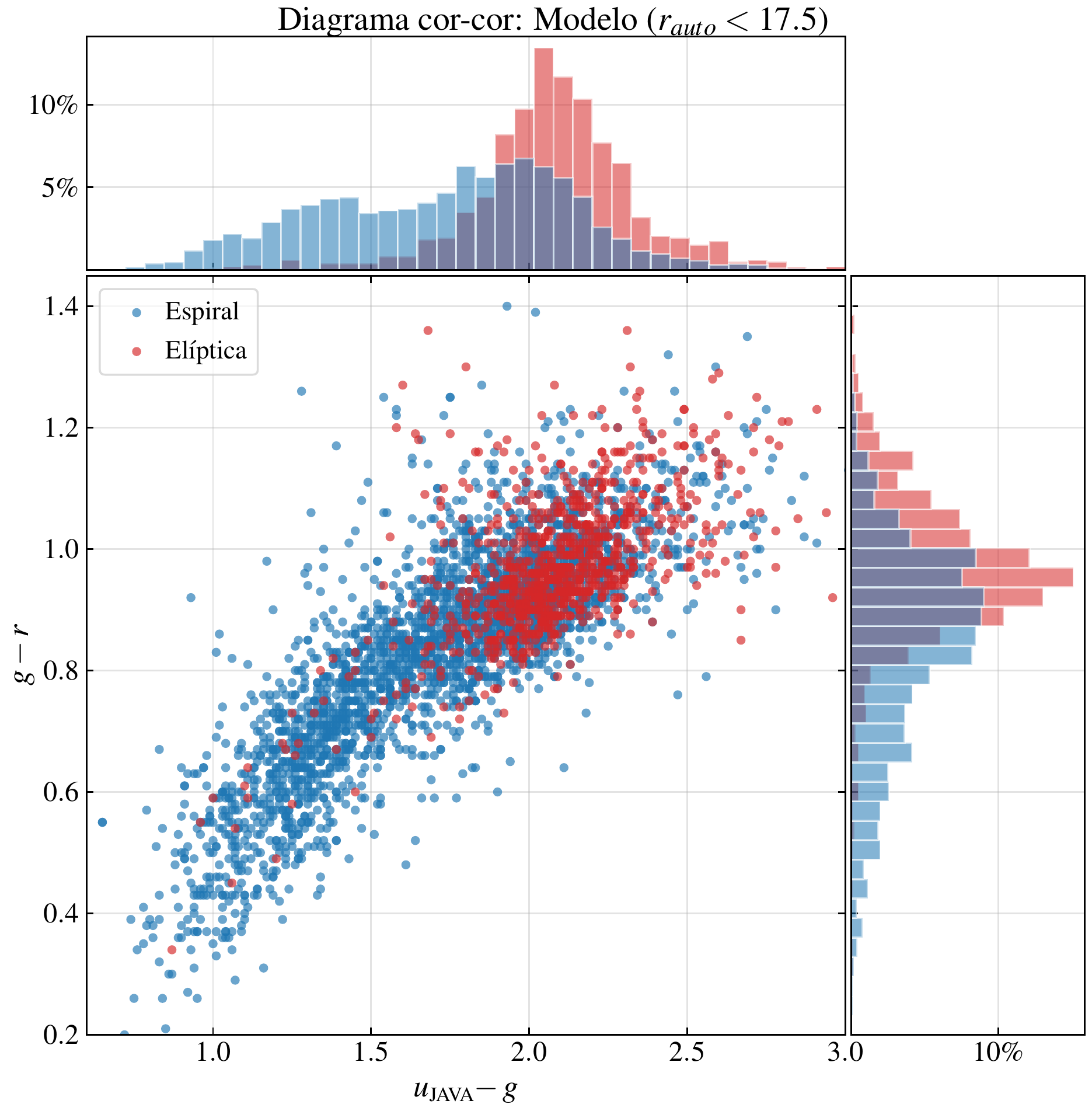}
  \end{subfigure}\\
  \begin{subfigure}{.5\linewidth}
    \includegraphics[width=\linewidth]{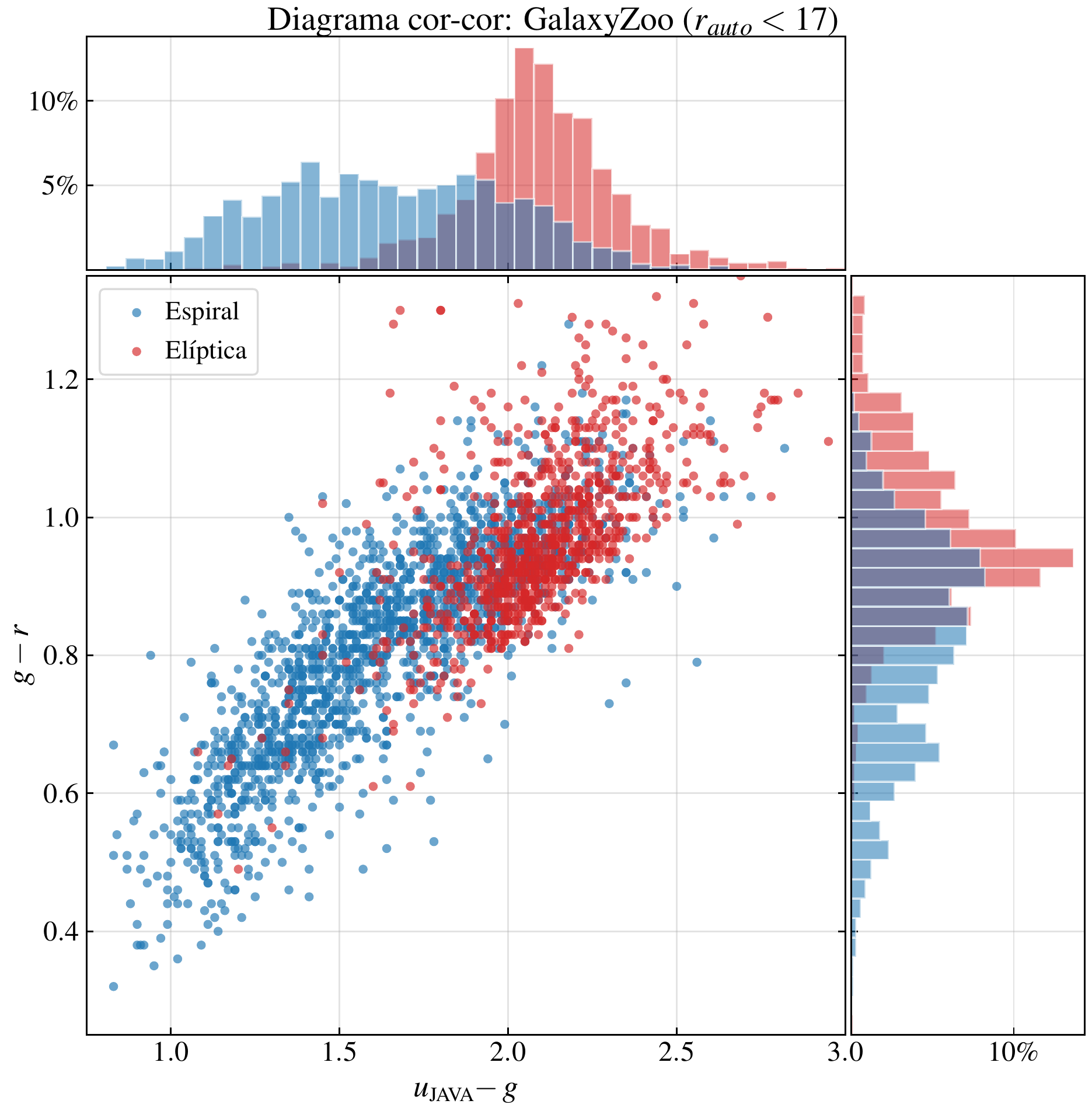}
  \end{subfigure}%
  \begin{subfigure}{.5\linewidth}
    \includegraphics[width=\linewidth]{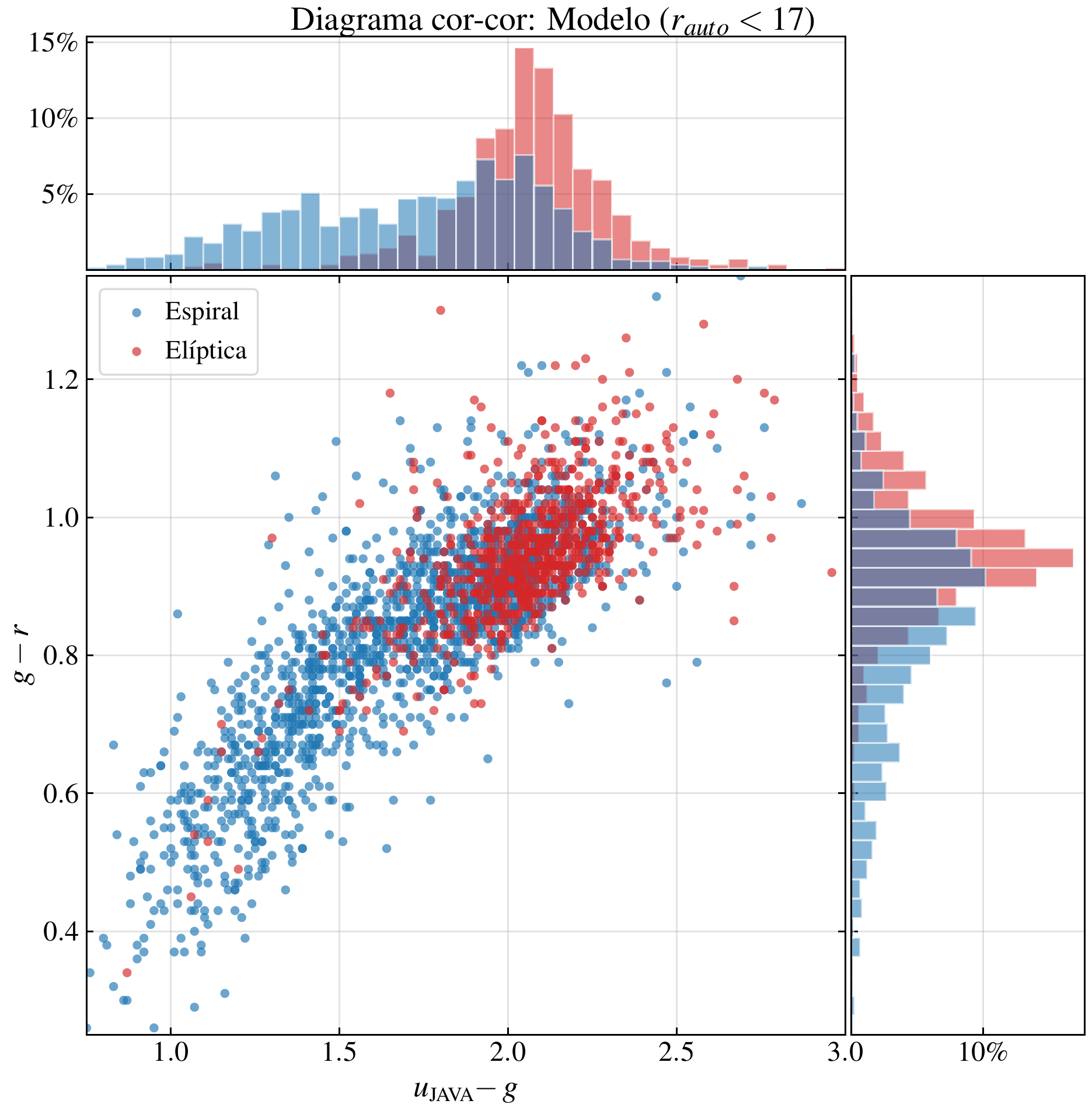}
  \end{subfigure}
  \caption{Diagrama cor-cor de $u_{JAVA} - g$ versus $g - r$. À esqueda, classificações visuais do GalaxyZoo nos conjuntos de treino, validação e teste e, à direita, classificações na amostra Blind feitas pelos modelos treinados com $r_{auto} < 17$ em cima e $r_{auto} < 17.5$ em baixo. Em azul, galáxias classificadas como Espirais e, em vermelho, galáxias classificadas como Elípticas.}%
  \label{fig:color-color}%
\end{figure*}

Existem várias métricas que ajudam a definir a capacidade de classificação de um modelo de \emph{deep learning}. Na Seção \ref{section:result-metricas} serão apresentadas as métricas utilizadas, suas expressões e o que elas avaliam. Na Seção \ref{section:result-treinamento}, será comparada a performance do modelo entre o conjunto de treinamento e de validação usando metricas quantitativas. Na Seção \ref{section:result-teste}, os modelos serão avaliados no conjunto de teste usando métricas quantitativas. E, na Seção \ref{section:result-blind}, será feita uma avaliação qualitativa dos modelos no conjunto blind.

\subsection{Métricas de avaliação quantitativa dos modelos}
\label{section:result-metricas}

As métricas aqui utilizadas baseiam-se no erro ou acerto na associação dos objetos às classes pelo modelo. Para relacionar a probabilidade calculada pelo modelo à classe, é definido um limiar de discriminação, que é a probabilidade mínima para que um exemplar pertença à uma classe. Para o cálculo das métricas, foi considerado um de limiar de 0.5.

A primeira métrica apresentada é a acurácia. Ela representa o número de objetos classificados corretamente em relação ao número total de objetos. Sua expressão é mostrada na equação \eqref{eq:acuracia}.

\begin{equation}
  \label{eq:acuracia}
  Acur\acute{a}cia = \frac{TP + TN}{TP + TN + FP + FN}
\end{equation}%
onde $TP$, $TN$, $FP$, e $FN$ são, respectivamente, a quantidade de Verdadeiro Positivo (\emph{True Positive}), Verdadeiro Negativo (\emph{True Negative}), Falso Positivo (\emph{False Positive}) e Falso Negativo (\emph{False Negative}).

Contudo, a acurácia nem sempre é uma métrica confiável para conjunto de dados desbalanceados, pois quanto maior a desproporção do número de objetos entre as classes, menor é o impacto das predições incorretas da classe em minoria no valor da acurácia, levando à uma avaliação superotimista do modelo. Para lidar com isso, usamos outras duas métricas: \emph{Precision} e \emph{Recall}, definidas nas equações \eqref{eq:precisao} e \eqref{eq:recall}, respectivamente.

\begin{equation}
  \label{eq:precisao}
  Precision = \frac{TP}{TP + FP}
\end{equation}

\begin{equation}
  \label{eq:recall}
  Recall = \frac{TP}{TP + FN}
\end{equation}

As equações \eqref{eq:precisao} e \eqref{eq:recall} mostram que $Precision$ tem o valor máximo na ausência de Falso Positivo e $Recall$ tem o valor máximo na ausência de Falso Negativo. Ao adotarmos a classe em minoria como positiva, temos uma avaliação que reflete a capacidade do modelo na tarefa mais difícil -- a classificação correta de objetos na classe com menor representação no conjuto de treinamento.
Para sumarizar estas duas métricas, usamos uma outra, chamada $F_1$\emph{-score}, que é a média harmônica entre $Precision$ e $Recall$, como definida na equação \eqref{eq:f1}.

\begin{equation}
  \label{eq:f1}
  F_1 = 2 \times \frac{Precision \times Recall}{Precision + Recall}
\end{equation}

\subsection{Performance dos modelos no treinamento}
\label{section:result-treinamento}

A Figura \ref{fig:train-metrics} mostra a avaliação do modelo em cada época de treinamento. Nos painéis superiores, criados a partir do conjunto de treinamento, notamos que, para a configuração escolhida, ambos os classificadores treinados executam a tarefa de minimizar a função de custo. Mas, para avaliar a capacidade de generalização da rede, ou seja, o potencial de reconhecimento de padrões em um caso real, é importante comparar os resultados com uma amostra diferente do treinamento. Os gráficos da acurácia no conjunto de validação, exibidos nos painéis inferiores, mostram que os classificadores conseguem bom desempenho em uma amostra não usada no treinamento.

Como visto na Seção \ref{section:divisao_conjunto_de_dados}, o modelo treinado no conjuto $r_{auto} < 17.5$ possui todos os objetos do conjunto $r_{auto} < 17$ mais objetos de magnitude entre 17 e 17.5. Sendo assim, a diferença de desempenho observada entre os modelos da direita e da esquerda mostram o impacto da inclusão destes objetos no treinamento. Este comportamento já era esperado, visto que são objetos difíceis de classificar até mesmo visualmente.

\subsection{Avaliação do modelo no conjunto de teste}
\label{section:result-teste}

A Tabela \ref{tab:best_hip} sumariza as melhores configurações obtidas para cada uma das arquiteturas testadas (conforme a Seção \ref{section:classificador}) e suas respectivas avaliações no conjunto de teste. Na parte inferior da tabela, foram incluídas as métricas definidas em \ref{section:result-metricas}, para cada classificador, com os melhores hiperparâmetros, como definidos em \ref{section:hyperparam}. Note que os resultados listados na tabela mostram a configuração dos classificadores que comporão o \emph{Ensemble}, com exceção dos Modelos E (EfficientNetB7) e F (VGG16), que não serão incluídos no \emph{Emsemble}, por não apresentarem resultados tão bons quanto os demais. A avaliação da rede VGG16 apresenta um alto desvio padrão para todas as métricas e a rede EfficientNetB7 não apresenta resultados compatíveis com o custo computacional.

\begin{table}[!h]
  \centering
  \caption{Avaliação dos modelos no conjunto de teste.}
  \label{tab:ensemble}
  \begin{tabular}{lrr}
    \toprule
    Métrica (\%) & $r_{auto} < 17$  & $r_{auto} < 17.5$ \\
    \midrule
    Acurácia     & $98.52 \pm 0.13$ & $93.48 \pm 0.12$  \\
    $F_1$-Score  & $98.52 \pm 0.13$ & $92.42 \pm 0.11$  \\
    ROC AUC      & $99.81 \pm 0.01$ & $98.79 \pm 0.01$  \\
    PR AUC       & $99.82 \pm 0.01$ & $98.94 \pm 0.01$  \\
    \bottomrule
  \end{tabular}
\end{table}

A Tabela \ref{tab:ensemble} mostra os resultados para o \emph{Ensemble}, obtido ao se combinar os modelos de A a D, como descrito na Seção \ref{section:meta-modelo}. Outra análise, mais robusta do que é mostrado nessa tabela, é avaliar os modelos sob diferentes limiares de discriminação, ao invés de sob apenas um. Para isto, usamos a Curva Característica de Operação do Receptor \emph{(Receiver Operating Characteristic -- ROC)} \cite{Hanley1982,Fawcett2006}, que é um gráfico de $FPR$ (False Positive Rate) versus $TPR$ (True Positive Rate). Usamos também a \emph{Curva Precision-Recall} (PR) que é um diagnóstico pontente para avaliar a classificação de modelos treinados a partir de conjuntos de dados desbanlanceados, como é o caso aqui.

A curva ROC pode ser visualizadas na Figura \ref{fig:roc}, onde resultados para os modelos individuais e para o Emsemble são comparados.
Note que a curva ROC é tão melhor quanto maior for a área abaixo desta (\emph{Area Under Curve -- AUC}), pois essa área representa o grau de separabilidade de um modelo. Na Figura \ref{fig:pr} mostramos a curva PR e o mesmo comentário vale aqui sobre o AUC. Fica claro ao analisar ambas as Figuras \ref{fig:roc} e \ref{fig:pr}, que os modelos são comparáveis entre si (as áreas sob as curvas são parecidas), e que os resultados do Emsemble são claramente melhores, AUC maiores, tanto para $r_{auto} < 17$ quanto para $r_{auto} < 17,5$. Outra conclusão ao se inspecionar as figuras em questão é que o Ensemble atinge resultados numericamente melhores no caso de classificação de objetos com $r_{auto} < 17$, do que com objetos com $r_{auto} < 17.5$.

\subsection{Avaliação qualitativa do modelo no conjunto blind: diagramas cor-cor}
\label{section:result-blind}

Um dos objetivos deste trabalho é obter classificações de galáxias elípticas e espirais, que não foram ainda classificadas (nas chamadas amostras \emph{blind}). Isto foi feito para uma amostra contendo 2536 galáxias com magnitude $r_{auto} < 17$ e 3951 galáxias com  magnitude $r_{auto} < 17.5$, com distribuições de magnitude e redshift compatíveis com as amostras de treinamento, como mostrado na Seção \ref{section:divisao_conjunto_de_dados}.

Uma maneira mais qualitativa de avaliar os resultados da classificação é pela análise do diagrama cor-cor das galáxias, comparando as nossas classificações com a do Galaxy Zoo. Nesta análise, é muito importante lidar com  amostras com distribuições de magnitude e redshift similares, como visto na Seção \ref{section:divisao_conjunto_de_dados}.

A Figura \ref{fig:color-color} mostra um diagrama de $u_{JAVA} - g$ versus $g - r$. Os painéis da esquerda mostram classificações visuais obtidas do GalaxyZoo nos conjuntos de treino, validação e teste enquanto que os painéis da direita mostram classificações feitas pelos nossos modelos descritos neste trabalho, no conjunto Blind. Como esta comparação é feita entre amostras diferentes de galáxias (embora com distribuição de brilho e redshift similares), não é esperado que os gráficos da esquerda sejam idênticos aos da direita, mas que tenham formas parecidas. Logo, o que se pretende mostrar é que a semelhança entre os diagramas da classificação humana com a classificação automática (usando nosso modelo) é um indicativo de uma classificação robusta.

\begin{figure*}[!ht]
  \centering
  \includegraphics[width=0.967\textwidth]{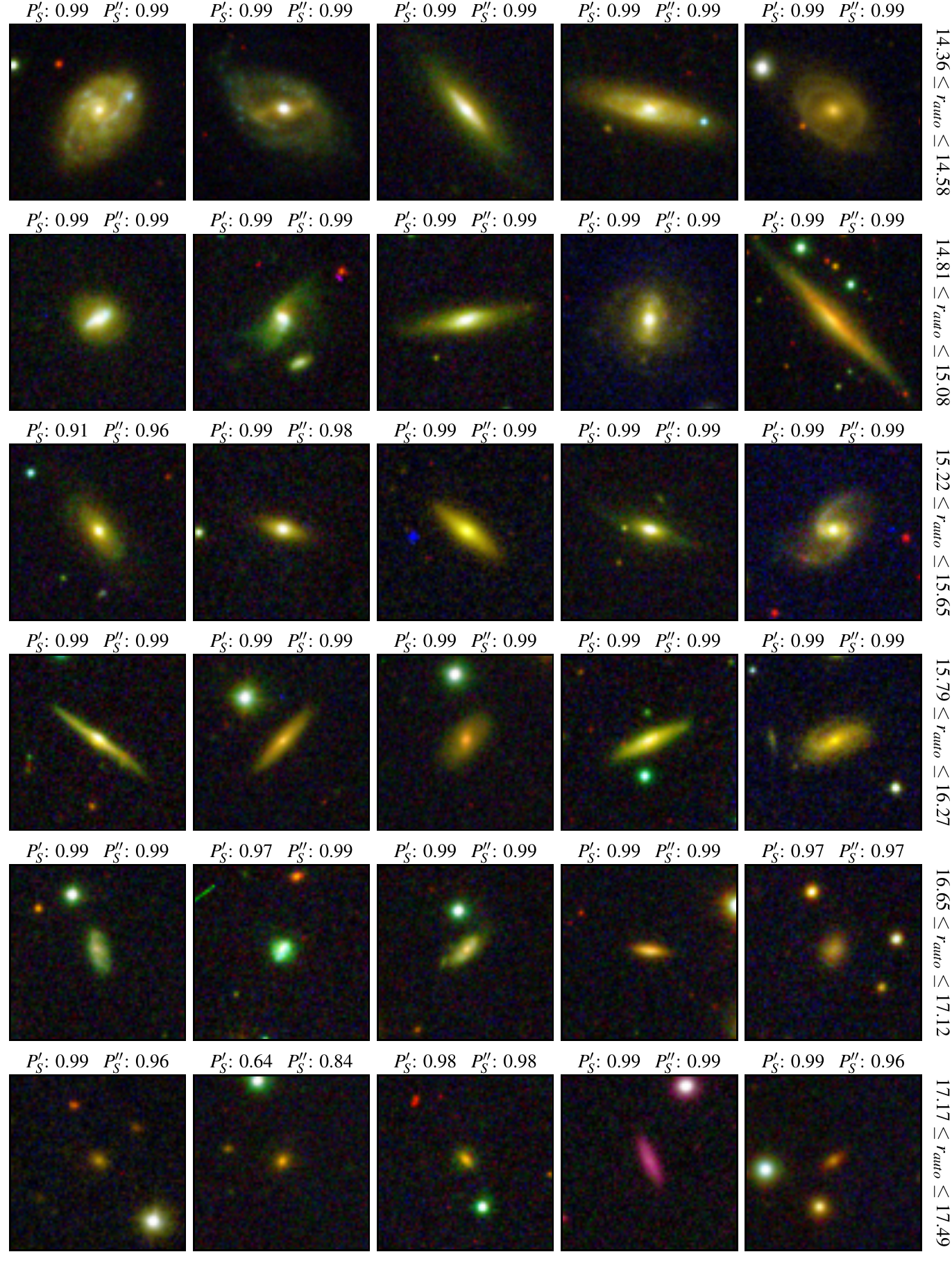}%
  \caption{Amostra aleatória com 30 galáxias do conjunto \emph{blind} com classificações dadas pelos modelos. Onde $P_S'$ e $P_S''$ representam a probabilidade da galáxia pertencer à classe espiral, inferida pelo modelo treinado com galáxias no intervalo de magnitude $r_{auto} < 17$ e $r_{auto} < 17.5$, respectivamente. As galáxias estão dispostas em ordem crescente de magnitude e o intervalo de $r_{auto}$ de cada linha é mostrado no lado direito.}%
  \label{fig:grid01}%
\end{figure*}

\begin{figure*}[!ht]
  \centering
  \includegraphics[width=0.967\textwidth]{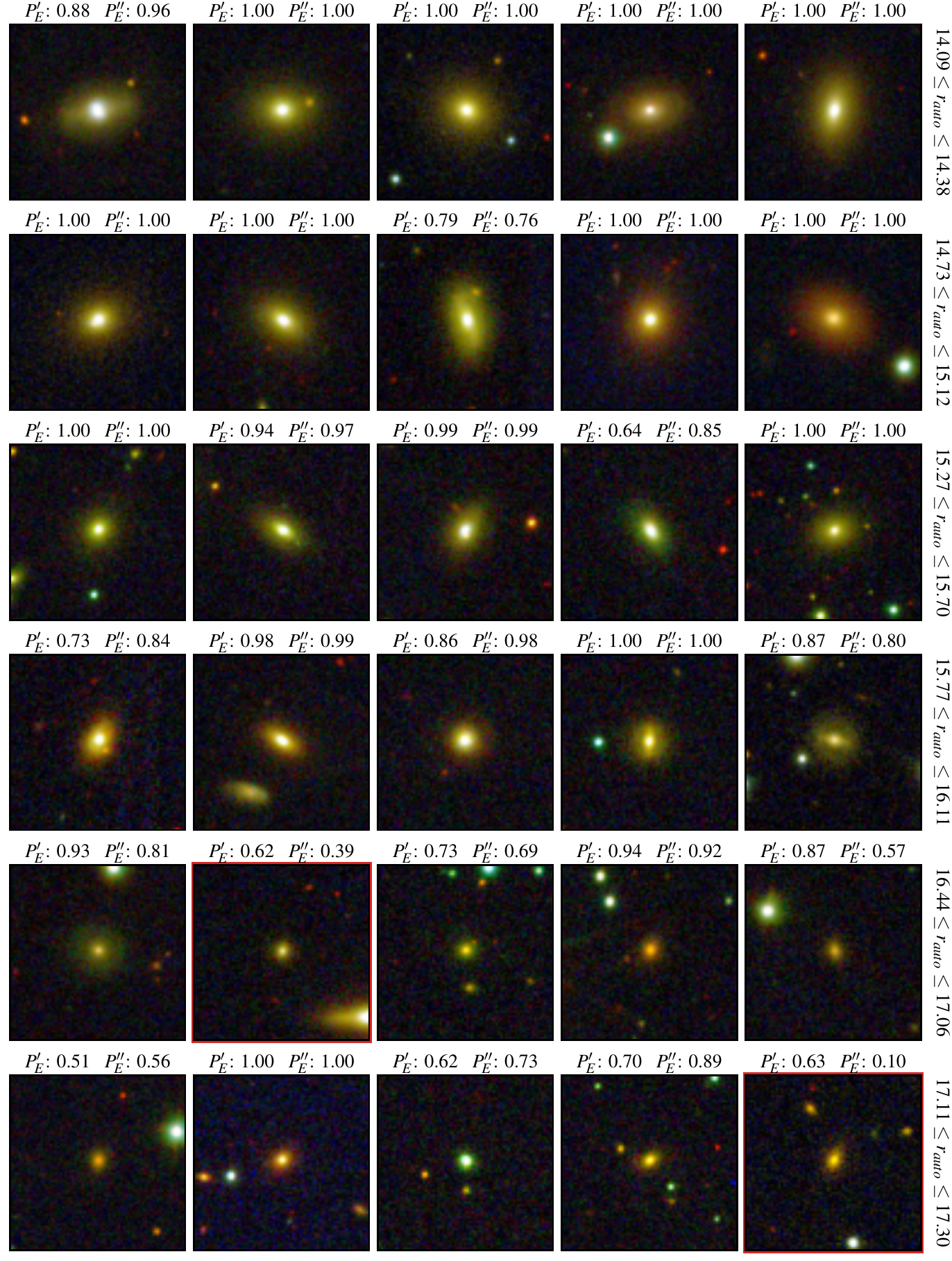}%
  \caption{Amostra aleatória com 30 galáxias do conjunto \emph{blind} com classificações dadas pelos modelos. Onde $P_E'$ e $P_E''$ representam a probabilidade da galáxia pertencer à classe elíptica, inferida pelo modelo treinado com galáxias no intervalo de magnitude $r_{auto} < 17$ e $r_{auto} < 17.5$, respectivamente. Painéis com bordas vermelhas assinalam para divergência na classificação dos modelos (para um limiar de 0.5).}%
  \label{fig:grid02}%
\end{figure*}

Um outro resultado mostrado nessas figuras é que as classificações são consistentes tanto para o modelo treinado usando galáxias com $r_{auto} < 17$ quanto para $r_{auto} < 17.5$. Aconselhamos que o modelo treinado em $r_{auto} < 17$ seja utilizado para classificação de galáxias até esta magnitude e o modelo $r_{auto} < 17.5$ seja utilizado em galáxias com magnitudes no intervalo de 17 a 17.5. As Figuras \ref{fig:grid01} e \ref{fig:grid02} mostram algumas das classificações dos modelos em amostras do conjunto blind.

\section{Discussão e Conclusão}

Em contraste com as abordagens de aprendizagem comuns, o que fizemos neste trabalho foi construir modelos a partir dos dados de treinamento, escolher entre os melhores modelos e combiná-los. O resultado principal deste trabalho é, então, a combinação das predições de várias redes com seus respectivos melhores hiperparâmetros e a constatação de que as métricas que avaliam os resultados mostram significativa melhoria quando usamos o método de emsemble, comparado com o uso de apenas um modelo. O objetivo final era o de obter maior acurácia na classificação de galáxias do mapeamento S-PLUS, e este objetivo é atingido através do método de emsemble aqui utilizado. O produto final é a classificação de galáxias no mapeamento S-PLUS em elípticas e espirais utilizando diversas redes. A morfologia das galáxias é uma propriedade fundamental necessária, por exemplo, para estudos de formação e evolução de galáxias.

Exploramos o uso de um \emph{Ensemble}, aumentando a acurária dos resultados e diminuindo a variância das predições e proporcionando melhores resultados em relação ao que é obtido com apenas o melhor classificador indidual. Outra novidade foi testar magnitudes até $17.5$, dado que um trabalho anterior \cite{bom2021} tinha classificado galáxias com magnitude até $17$ com um classificador apenas.

Comparando este trabalho com o de \cite{bom2021}, para o limite de magnitude $17$, notamos resultados bastante similares, sendo que aquele trabalho levou em consideração imagens FITS enquanto que este trabalho utilizou imagens RGB, obtidas utilizando o \emph{software} Trilogy~\cite{coe2012clash}. Do ponto de vista computacional, muitas vezes a utilização de imagens RGB é a única opção, pois as imagens em formato FITS podem ocupar um maior espaço na memória de GPU, aumentando o tempo de processamento. Logo, foi importante mostrar aqui que as imagens RGB dão resultados similares quando o Ensemble é utilizado.

Um ponto muito importante neste trabalho é o pré-processamento. Este incluiu padronização, escalonamento e/ou normalização, como descrito na Seção \ref{section:preparacao}. O importante é que se faça o procedimento adequado para cada arquitetura. No caso específico deste trabalho, esta foi a maneira que apresentou o melhor desempenho quando combinado com a inicialização dos pesos com pré-treino na base de dados \emph{ImageNet}. Um resultado similar é mostrado no trabalho de \cite{bom2021}, em que as redes treinadas usando o pré-treino \emph{ImageNet} ultrapassaram a performance das redes treinadas sem pré-treino, mesmo com um pré-processamento de dados diferente do que foi feito aqui.

Outro ponto importante, que foi testado neste trabalho, é que, apesar de, tanto a escolha da arquitetura quanto dos hiperparâmetros da rede serem terem impacto nos resultados, em trabalhos onde apenas uma rede é utilizada, no nosso caso este impacto é minimizado pela combinação dos classificadores. Desta forma, as diferenças dos resultados causadas por variações nos hiperparâmetros de cada rede são atenuadas.

Este trabalho mostrou a melhoria dos resultados usando método de \emph{Ensemble}. No futuro, este modelo pode ser usado para classificação de galáxias do levantamento S-PLUS.

\section{Disponibilidade dos dados}
Nós disponibilizamos publicamente os catálogos de classificação bem como os modelos de \emph{Deep Learning} na página \url{https://natanael.net}.

\section{Agradecimentos}
Os modelos foram implementados usando vários projetos \emph{open-source}, como a linguagem de programação Python \cite{python}, o Trilogy \cite{trilogy}, as bibliotecas de \emph{deep learning} Tensorflow \cite{tensorflow} e Keras \cite{keras} e outras bibliotecas de computação científica \cite{astropy, sklearn, skimg, numpy, scipy, ipython, pandas, matplotlib}. Este projeto também fez uso de serviços online, como o SkyServer\footnote{\url{http://skyserver.sdss.org/dr16/en/home.aspx}} e o S-PLUS Cloud\footnote{\url{https://splus.cloud}}, para acesso de imagens e catálogos astronômicos, e o Google Colab\footnote{\url{https://colab.research.google.com}}, para treinamento dos modelos usando GPU's gratuitamente. Gostaríamos de agradecer os colegas Arianna Cortesi, Geferson Lucatelli, Erik Lima, além do grupo de trabalho de Morfologia, pelas importantes contribuições e discussões sobre o assunto.

\clearpage
\bibliographystyle{unsrt}
\bibliography{bib}
\end{document}